\documentclass[10pt,twocolumn,superscriptaddress,floats,showpacs,nobalancelastpage,longbibliography]{revtex4-1}

\usepackage{array,longtable}
\newcolumntype{L}{>{\tiny $}p{0.33\columnwidth}<{$}}
\newcolumntype{M}{>{\scriptsize $}p{0.33\columnwidth}<{$}}
\newcolumntype{N}{>{\scriptsize $}p{0.43\columnwidth}<{$}}
\usepackage{amsmath}
\usepackage{amssymb}
\usepackage{amsfonts}
\usepackage{graphicx}
\usepackage{hyperref}

\usepackage{float}
\usepackage{graphics}
\usepackage[caption=false]{subfig}

\allowdisplaybreaks[1]

\newif\ifhyper
\hypertrue
\ifhyper
\hypersetup{
   citecolor = {green},
   colorlinks = {true}, 
   urlcolor = {blue} 
}
\fi

\begin{document}

\title{Discrete lattice symmetry breaking in a two-dimensional frustrated spin-1 Heisenberg model}

\author{Ji-Yao Chen}
\affiliation{Laboratoire de Physique Th\'eorique, IRSAMC, Universit\'e de Toulouse, CNRS, UPS, 31062 Toulouse, France}
\author{Sylvain Capponi}
\affiliation{Laboratoire de Physique Th\'eorique, IRSAMC, Universit\'e de Toulouse, CNRS, UPS, 31062 Toulouse, France}
\author{Didier Poilblanc}
\affiliation{Laboratoire de Physique Th\'eorique, IRSAMC, Universit\'e de Toulouse, CNRS, UPS, 31062 Toulouse, France}

\date{\today}

\begin{abstract}

Spontaneous discrete symmetry breaking can be described in the framework of Projected Entangled Pair States
(PEPS) by linearly superposing local tensors belonging to two (or more) symmetry classes of tensors. This is illustrated in the case of a frustrated spin-1 Heisenberg model on the square lattice, which hosts a nematic spin liquid spontaneously breaking lattice $\pi/2$-rotation symmetry. A superposition of SU(2)-symmetric PEPS tensors belonging to two irreducible representations of the lattice point group is shown to capture accurately the properties of the nematic phase, as shown from a comparison to Exact Diagonalisations and Density Matrix Renormalization Group results. 
 
\end{abstract}
\maketitle


{\it Introduction and model --} 
The Ginzburg-Landau (GL) paradigm offers a general and elegant framework to describe low-energy phases in condensed matter systems in terms of spontaneous symmetry breaking. For example, magnetic ordering in quantum spin systems can be viewed as a spontaneous breaking of the underlying spin rotation (SU(2)) symmetry. Electronic charge density waves or (spin) valence bond crystals spontaneously break the
underlying lattice translation symmetry. More interestingly,  in the nematic electronic phases of some  pnictides~\cite{Chuang2010,Fradkin2010} (a class of iron-based high-Tc superconductors) orientational order sets up
that resembles -- to some extent -- classical liquid crystals of molecules~\cite{Ermakov2016}. In the electronic system, the lattice (discrete) point group symmetry is spontaneously broken, a scenario fitting well into the GL scheme.
E.g. on the two-dimensional (2D) square lattice, of point group $C_{4v}$, the horizontal ($x$) and vertical ($y$) axis become non-equivalent, so that transport or correlations along these two directions become different. 

Our aim here is to investigate such a phenomenon in simple quantum spin magnets. 
In such a case, the nematic phase can be viewed as a melted (thermal~\cite{Weber2003, Schecter2017} or quantum~\cite{Fradkin2010}) magnetic stripe phase (of magnetic wave vector ${\bf q}=(\pi,0)$ or $(0,\pi)$) where the SU(2) spin symmetry has been restored while orientational order (spontaneous breaking of the 90-degree lattice rotation) still persists. Therefore, nematic phases could potentially appear in the immediate proximity of magnetic stripe phases. 
Alternatively, nematic phases can also emerge from the N\'eel phase via a proliferation of
monopoles~\cite{Wang2015}.  Monopoles carry Berry phases~\cite{Haldane1988} implying
that, in the phase where the monopoles proliferate with spin $S=1$ (or odd-integer spin) there is nematic order~\cite{Read1989}.

To be more specific, let us consider the frustrated spin-$1$ Heisenberg model on the two-dimensional square lattice,
\begin{align}
H&=J_1\sum_{\big< i,j\big>} {\bf S}_i\cdot{\bf S}_j
+ J_2\sum_{\big<\big<k,l\big>\big>} {\bf S}_k\cdot{\bf S}_l\nonumber \\
&+ K_1\sum_{\big< i,j\big>} ({\bf S}_i\cdot{\bf S}_j)^2 ,
\label{EQ:model}
\end{align}
where the first and third sums are taken over nearest-neighbor (NN) bonds and the second sum runs over next-nearest-neighbor (NNN) bonds. 
For simplicity, we set the NN and NNN bilinear couplings to $J_1=1$ and $J_2=0.54$, corresponding approximately to the maximally frustrated regime~\cite{Jiang2009}, and vary the biquadratic coupling $K_1$. As shown in Ref.~\onlinecite{Jiang2009}, the $J_1$-$J_2$ model (with spatial anisotropy of the $J_1$ coupling) may host a quantum-disordered phase, in a narrow region of parameter space, between a N\'eel phase and a (magnetic) stripe phase.
We argue in this paper, that the quantum-disorder phase spontaneously breaks lattice rotation symmetry, being of nematic character.
For our choice of the bilinear couplings the nematic phase is stable in the range $0\lesssim K_1 \lesssim 0.15$.

Our strategy to explore the physics of the above model is to combine different numerical techniques such as
Exact Diagonalisations (ED),  Density Matrix Renormalization Group (DMRG) and 
tensor network methods~\cite{Cirac2009b,Cirac2012a,Orus2013,Schuch2013b,Orus2014}.
In particular, we shall focus on SU(2)-symmetric Projected Entangled Pair States (PEPS) to describe the nematic spin liquid phase. 
More precisely, we construct an explicit PEPS wave function that provides a faithful representation of the symmetry-broken non-magnetic state directly in the thermodynamic limit. 
In contrast to usual PEPS calculations, which approach the ground state of the model via imaginary time evolution (starting from some initial random state), we use a more elegant framework. As is commonly known, the manifold of PEPS is not simply connected, but has many different phases, 
which makes the search for spin liquid phases in frustrated models challenging.
It is therefore appropriate, following the pioneering work in Ref.~\cite{Jiang2015},  to study each class separately -- with a clear understanding of the physical nature of the variational wave function --
using a variational optimization scheme. 

{\it Lanczos exact diagonalizations --}
As a preliminary study to explore the physics of Hamiltonian~\eqref{EQ:model},
we have performed Lanczos exact diagonalisations (ED) of 
finite periodic $4\times 4$ and $\sqrt{20}\times \sqrt{20}$ clusters (topologically equivalent to a torus). 
``Level spectroscopy'' is a useful tool to get insights on the nature of the ($T=0$) quantum phase. 
In genuine magnetically-ordered phases (spontaneously breaking the continuous SU(2) spin rotation symmetry) one expects low-energy triplets (so called Anderson's tower of states) collapsing onto the (singlet) ground state (GS) in the thermodynamic limit. In contrast, in the case of a ``spin liquid'' only breaking a discrete space group symmetry, one expects a finite number of quasi-degenerate GS separated from the rest of the spectrum by a finite gap. 
In addition, the quantum numbers associated to the quasi-degenerate GS give insights on the precise nature of the symmetry-broken phase. 
\begin{figure}[htbp]
\begin{center}
\includegraphics[width=\columnwidth,angle=0]{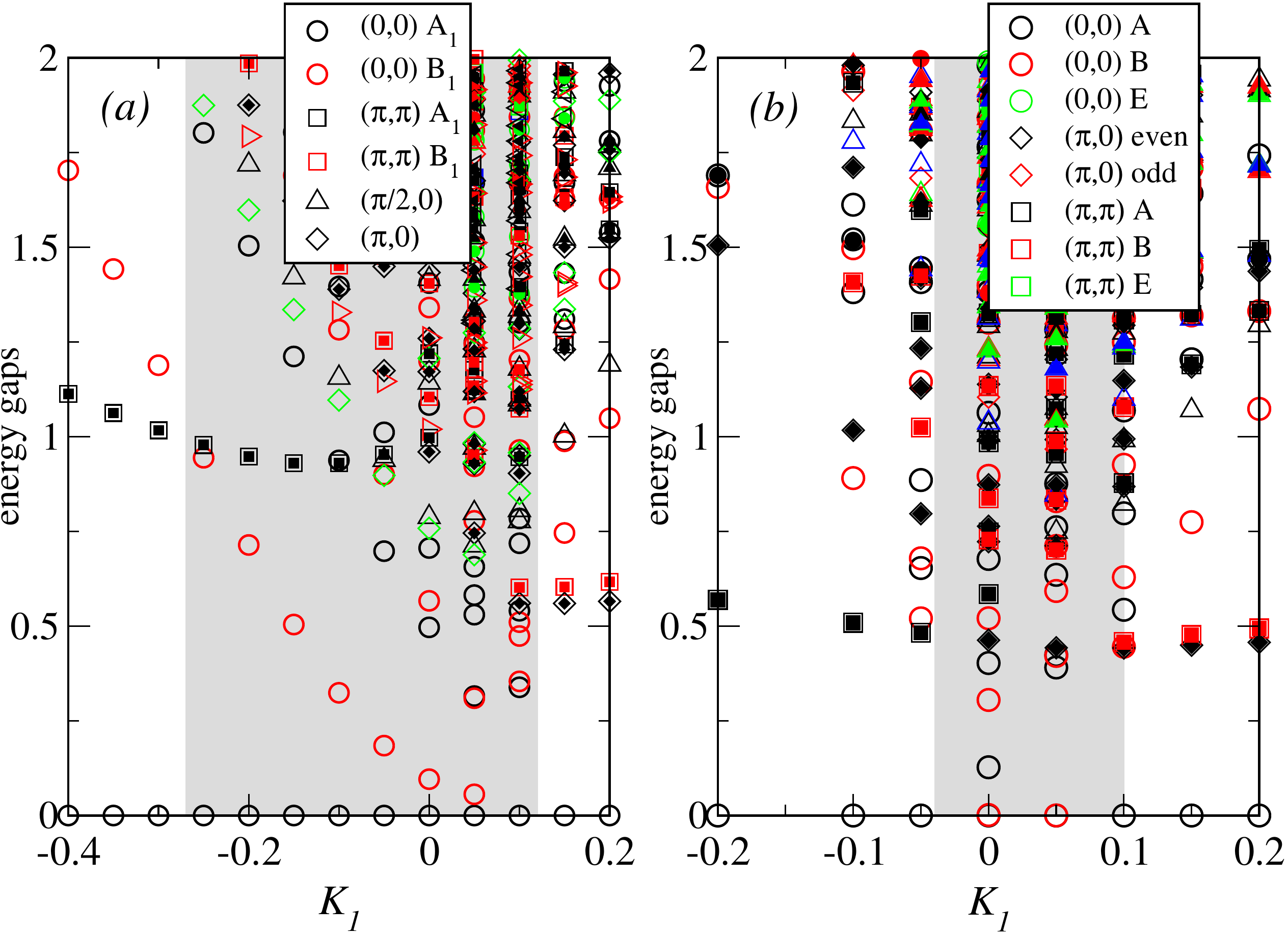}
\caption{
[Color online] Low-energy spectra of the Heisenberg Hamiltonian~\eqref{EQ:model} vs 
$K_1$ obtained by Lanczos ED of periodic $4\times 4$ (a) and $\sqrt{20}\times \sqrt{20}$ (b) clusters. 
The different symbols (colors)
correspond to different momenta (IRREP of the lattice point group). Open (filled) symbols correspond to singlet (triplet) states. The shaded / white regions are characterized by the singlet / triplet nature of the first excitation above the GS, suggesting possible nematic / magnetic phases (see text).
 }
\label{FIG:ED}
\end{center}
\end{figure}

Fig.~\ref{FIG:ED} shows a narrow region of the parameter $K_1$ where the two lowest eigenstates are 
momentum ${\bf q}=(0,0)$ singlets
belonging to two different $A$ and $B$ irreducible representations (IRREPs) of the point group. Note that, 
although the 20-site cluster does not possess all the square lattice symmetries (its point group is $C_4$ instead of $C_{4v}$ for the 16-site cluster and the infinite lattice) it nevertheless possesses the 90-degree rotation that
enables to distinguish between the $A$ and $B$ IRREPs.  In particular, for $K_1=0.05$ the energy separation between the two states is very small (even not visible in Fig.~\ref{FIG:ED}(b)) and a significant gap appears above. This low-energy spectrum is typical of a lattice nematic phase with non-equivalent correlations in the $x$- and $y$-axis directions. In the neighboring left and right regions the first excitation is a triplet state, suggesting the occurrence of magnetic phases. The momentum
${\bf q}=(\pi,\pi)$ (${\bf q}=(\pi,0)$ and $(0,\pi)$) of the first triplet excited state is compatible with 
N\'eel (stripe) magnetic order. Note however, that the extension of the supposed nematic spin liquid
phase strongly depends on the cluster, so that complementary methods are required to ascertain its stability.
\begin{figure}[!htbp]
\begin{center}
\includegraphics[width=\columnwidth,angle=0]{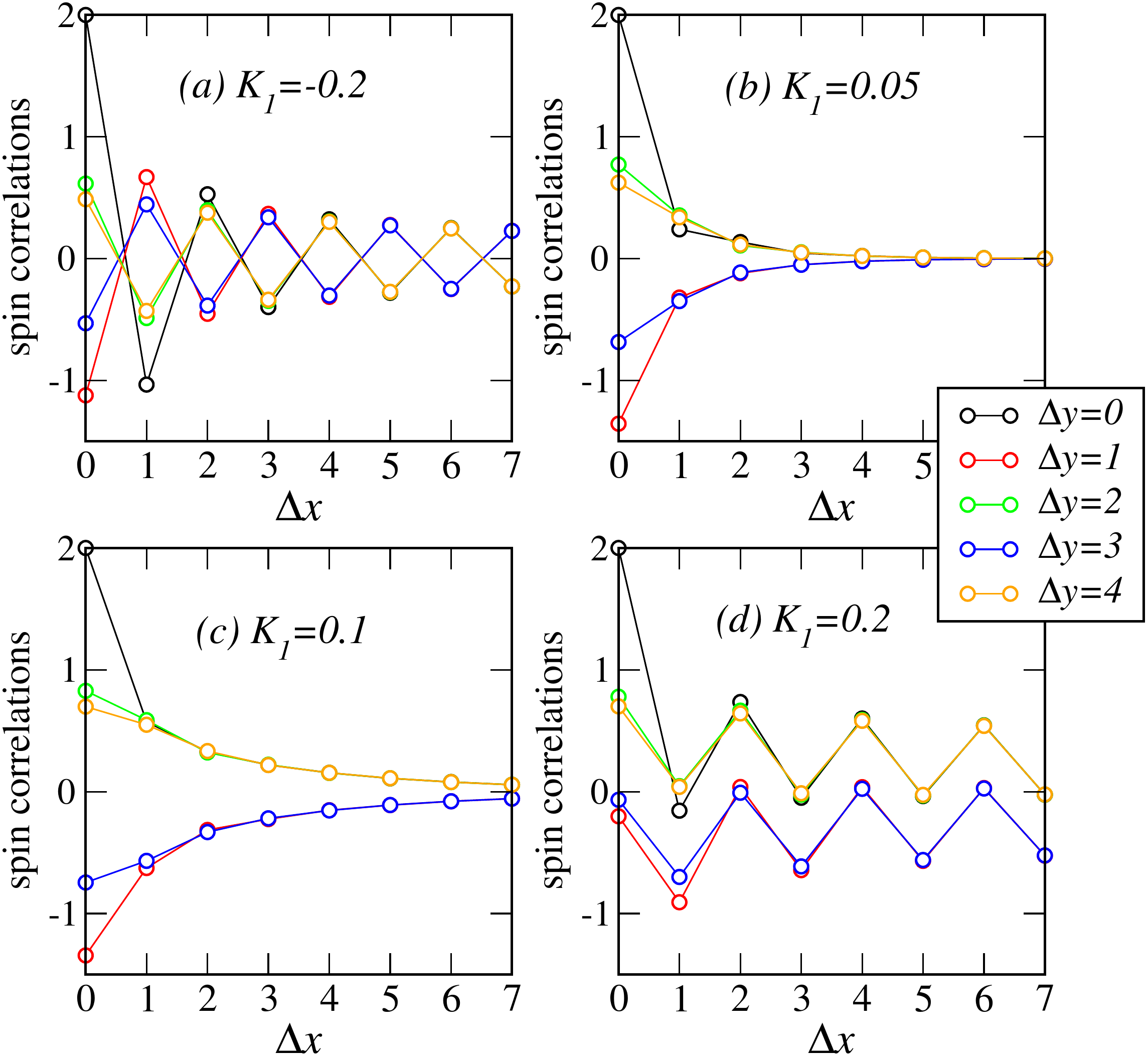}
\caption{
[Color online] DMRG results for spin-spin correlations (\ref{eq:corr}) at distance $(\Delta x,\Delta y)$ obtained on a $16\times 8$ cylinder for various $K_1$ indicated on the plot. Data are averaged in the middle $8\times 8$ region to reduce finite-size effects. 
 }
\label{fig:dmrg_corr}
\end{center}
\end{figure}

{\it DMRG phase diagram --} 
To better characterize the three phases at play, we have performed 
DMRG simulations on $L\times W$ cylinder, using typically $L=2W$ (with periodic boundary conditions in the shorter direction).  In order to characterize the putative phases, we have measured their respective order parameters. For nematicity, we directly measure the $x$ and $y$ bilinear bond energies in the bulk of the system~\footnote{Note that we get qualitatively similar results if we measure the full bond energies including the biquadratic term.} and define an order parameter as 
\begin{equation}\label{eq:nematic}
OP = | \langle {\bf S}_{x,y}\cdot{\bf S}_{x+1,y} - {\bf S}_{x,y}\cdot{\bf S}_{x,y+1} \rangle |.
\end{equation}
Note that since our cylinder explicitly breaks lattice rotation symmetry, we can obtain a finite value even on a finite cluster. But we do need to perform finite-size scaling to check its thermodynamic limit value, see below and Appendix~\ref{app:dmrg}. 
For magnetic phases (N\'eel or stripe), since our finite cluster cannot break SU(2) symmetry, we need to compute relevant spin-spin correlations:
\begin{equation}\label{eq:corr}
C(\Delta x,\Delta y) = \langle {\bf S}_{x,y}\cdot{\bf S}_{x+\Delta x,y+\Delta y}\rangle.
\end{equation}
In Fig.~\ref{fig:dmrg_corr}, we plot some examples of real-space spin-spin correlations for various $K_1$. It is already apparent that the modulation changes from $(\pi,\pi)$ wavevector for negative $K_1$ values to $(0,\pi)$ for positive ones. Note also that, since we are only using U(1) symmetry, our DMRG simulation can end up in a state with a finite local $\langle S_i^z\rangle$ (for instance, we measure $|\langle S_i^z\rangle| \sim 0.5$ when $K_1=\pm 0.2$), which suggests long-range magnetic order in the thermodynamic limit. Note however that, for intermediate values $K_1=0.05$ and $K_1=0.1$, there are no oscillations and correlations 
become very small at the largest available distances. 

In order to be more quantitative, we have chosen to measure all spin correlations within a $W\times W$ subset in the center of each $2W\times W$ cylinder in order to compute the structure factors $m^2(\pi,\pi)$ and $m^2(\pi,0)$ respectively: 
\begin{equation}
\label{eq:m2}
m^2({\bf q}) = \frac{1}{N_s} \sum_{i,j \in {\cal C}} \langle {\bf S}_i \cdot  {\bf S}_j\rangle e^{i {\bf q}\cdot {\bf r}},
\end{equation}
where the sum runs over all sites $(i,j)$ within the central part with $N_s=W^2$ sites and ${\bf r}$ is the relative distance between $i$ and $j$. While a cylinder geometry favors a modulation with wavevector $(0,\pi)$ for positive $K_1$ (see Fig.~\ref{fig:dmrg_corr}d), we rather consider here the structure factor at ${\bf q}=(\pi,0)$ that should also converge to the square of the order parameter in the thermodynamic limit and has less finite-size effects, see Appendix~\ref{app:dmrg}. 
\begin{figure}[!htbp]
\begin{center}
\includegraphics[width=\columnwidth,angle=0]{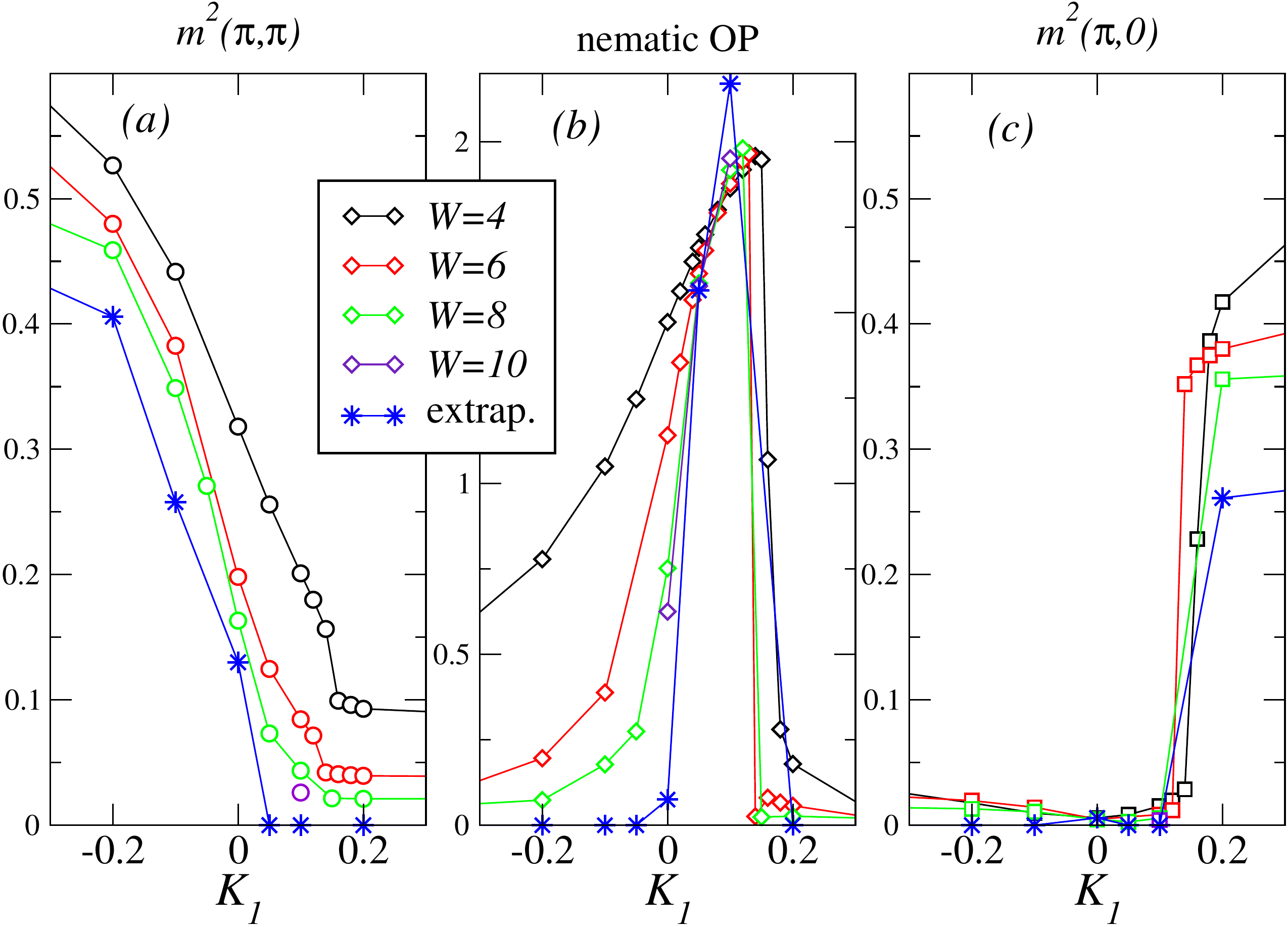}
\caption{
[Color online] DMRG results vs $K_1$ obtained on various $2W\times W$ cylinders (with open boundary conditions in the long direction, see text): (a) Square of the
staggered magnetization (\ref{eq:m2}); (b) nematic order parameter (\ref{eq:nematic}); (c) square of the ${\bf q}=(\pi,0)$ magnetic structure factor (\ref{eq:m2}). Extrapolation leads to a phase diagram with three successive phases: N\'eel, nematic and magnetic stripe phase respectively when increasing $K_1$ parameter. 
 }
\label{fig:dmrg}
\end{center}
\end{figure}

In Fig.~\ref{fig:dmrg}, we summarize our DMRG results plotting, side by side, the order parameters of the three phases as a function of $K_1$. Our data are compatible with the existence and stability of three phases: two magnetically ordered ones (N\'eel for $K_1\lesssim 0$, stripe for $K_1 \gtrsim 0.15$) and a nonmagnetic nematic phase in between~\footnote{Note that, strictly speaking, the nematic order parameter may not identically vanish in the stripe phase.}. Moreover, we observe rather sharp transitions (especially at positive $K_1$) so that presumably both quantum phase transitions are first order, as expected from a standard Ginzburg-Landau analysis.

\begin{figure}[htb]
\centering
	\begin{minipage}[c]{0.12\textwidth}
	\centering
	\subfloat[]{\includegraphics[width=16mm, height=16mm]{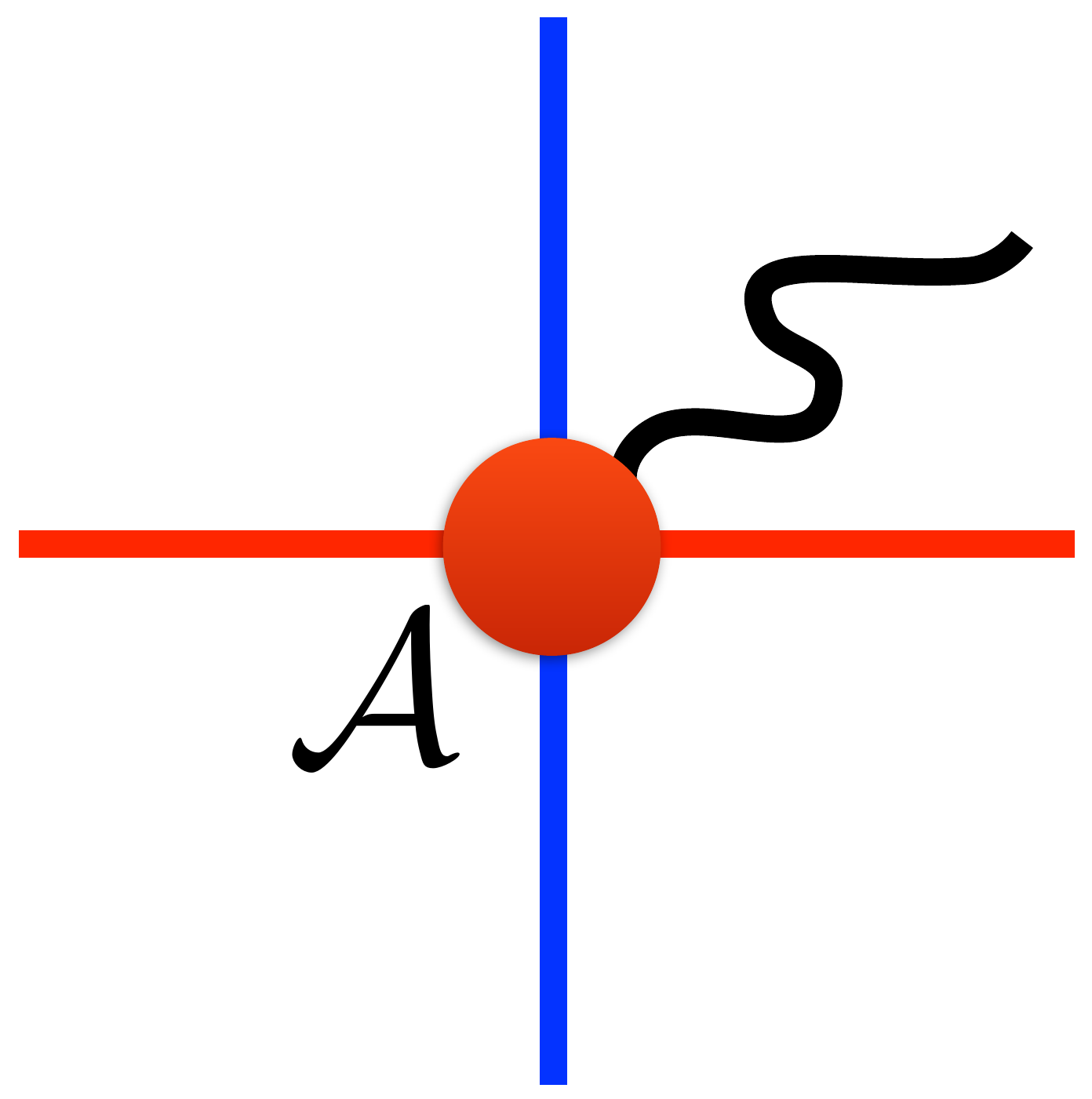}}\\
	\subfloat[]{\includegraphics[width=16mm, height=16mm]{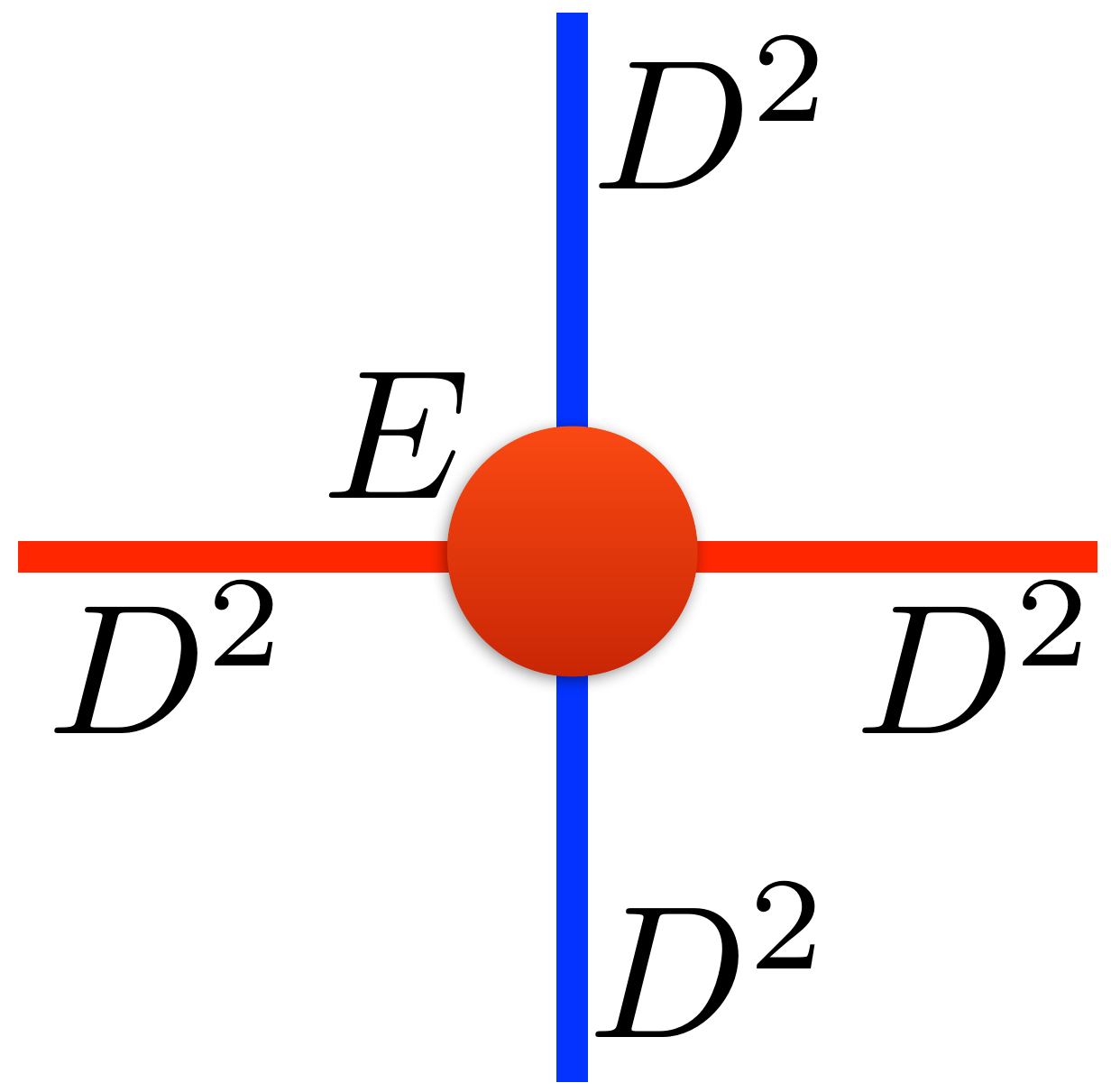}}
	\end{minipage}
	\begin{minipage}[]{0.35\textwidth}
	\centering
	\subfloat[]{\includegraphics[width=43mm, height=41.5mm]{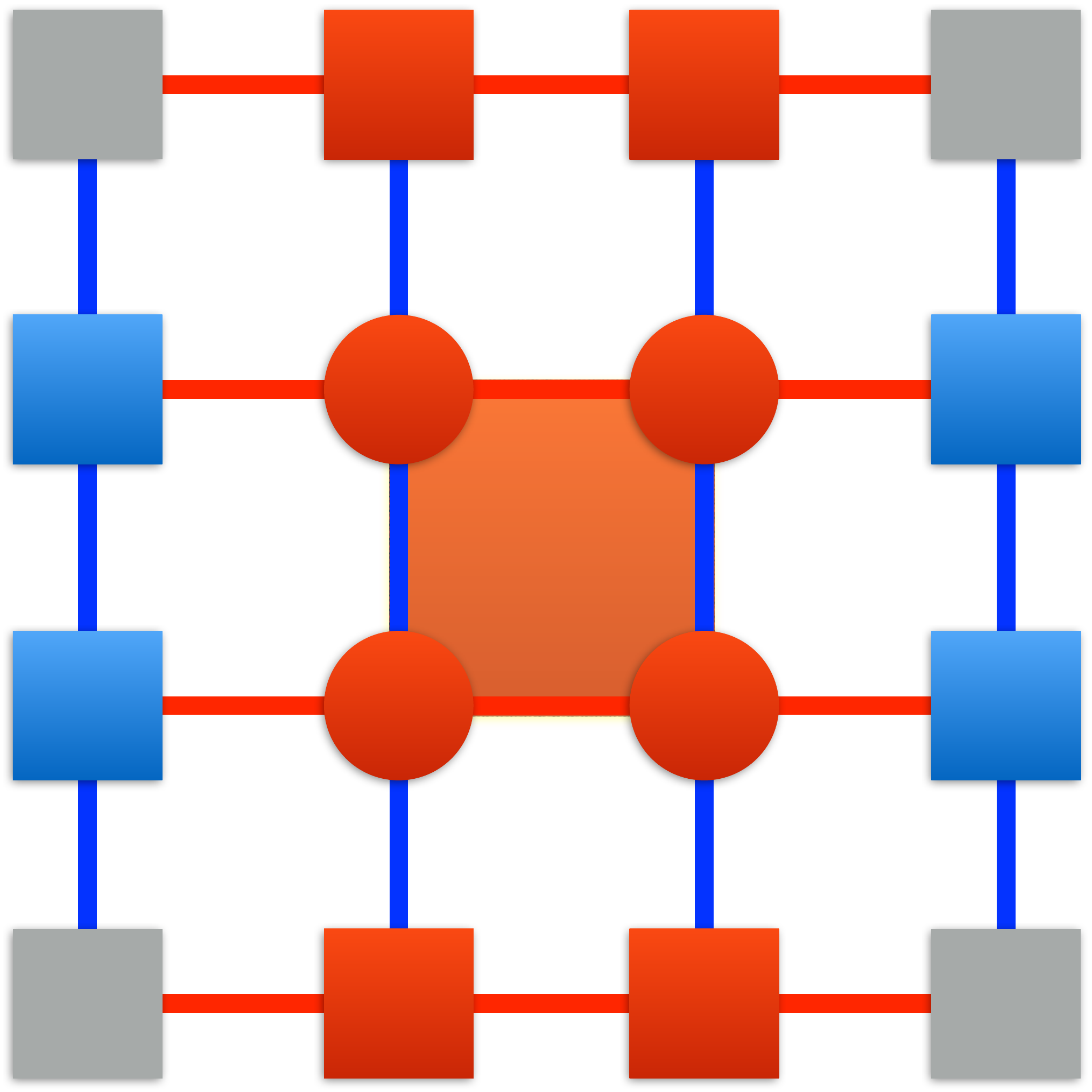}}
	\end{minipage}\\
	\subfloat[]{\includegraphics[width=18mm, height=16mm]{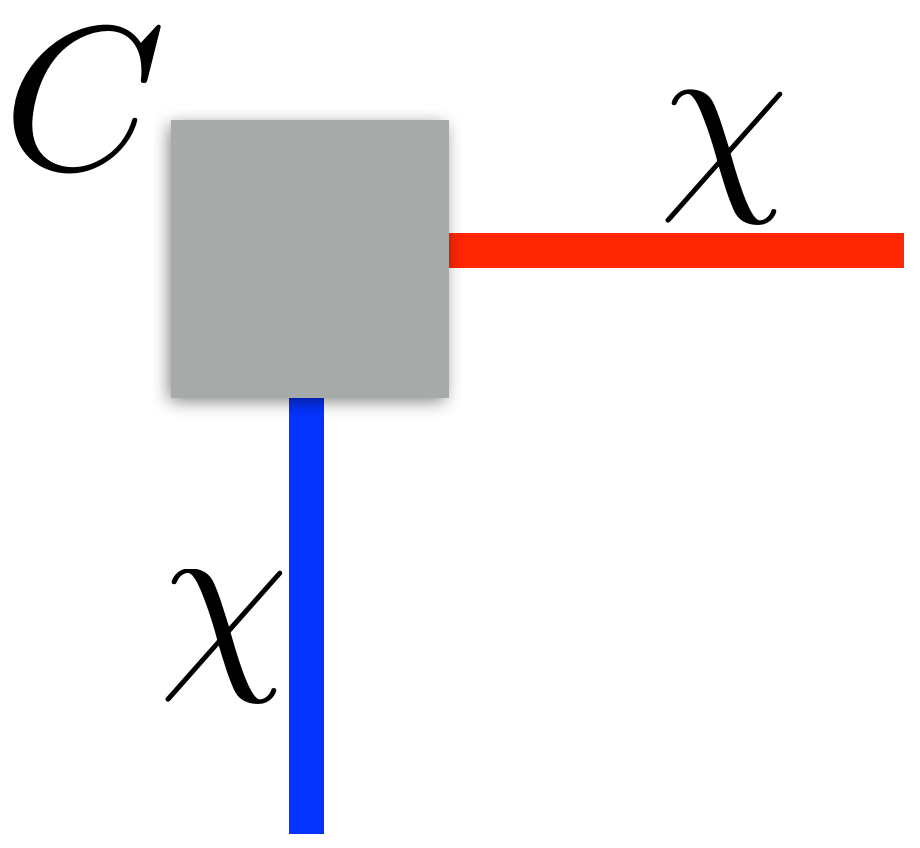}}\hfill
	\subfloat[]{\includegraphics[width=24mm, height=17mm]{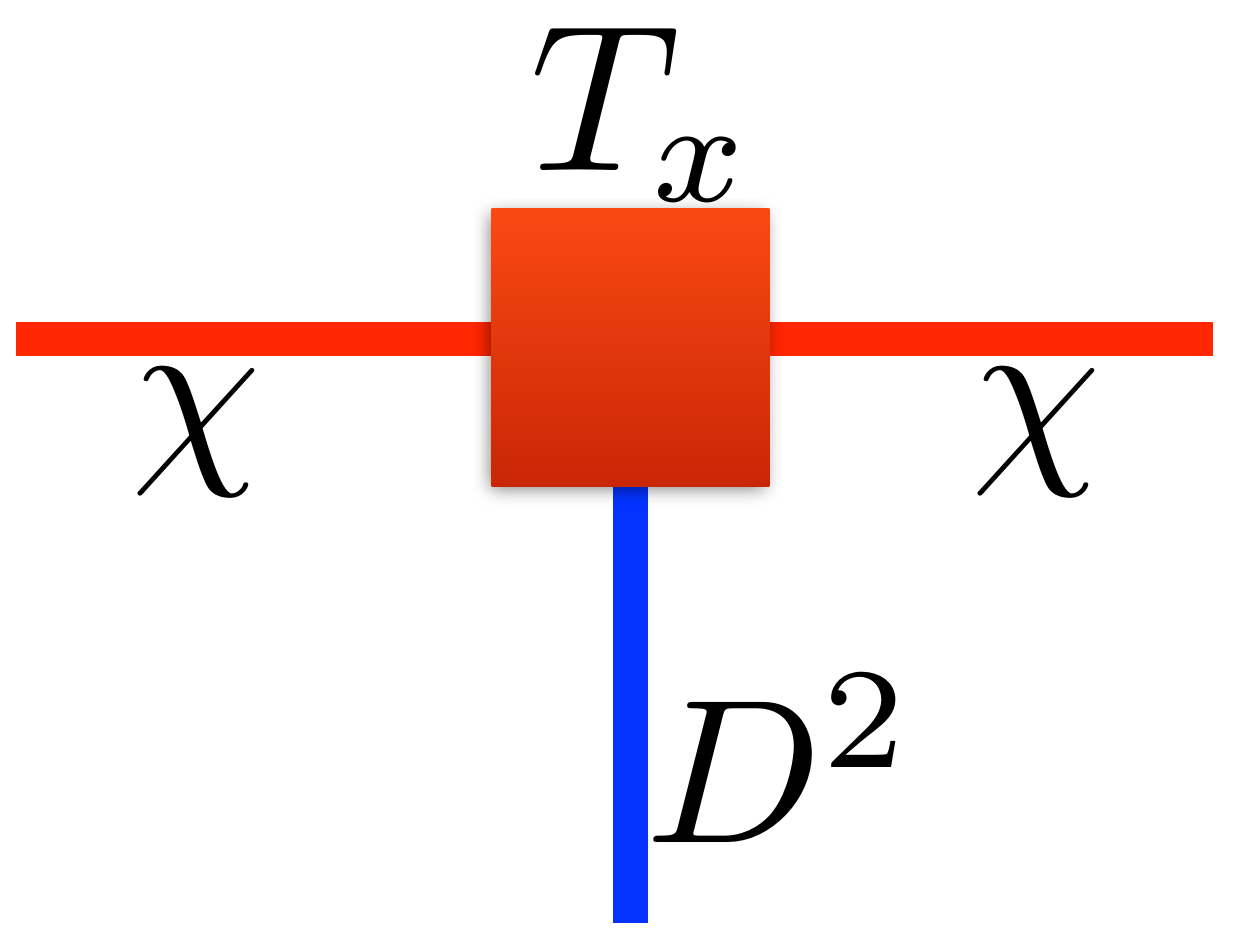}}\hfill
	\subfloat[]{\includegraphics[width=19.5mm, height=21mm]{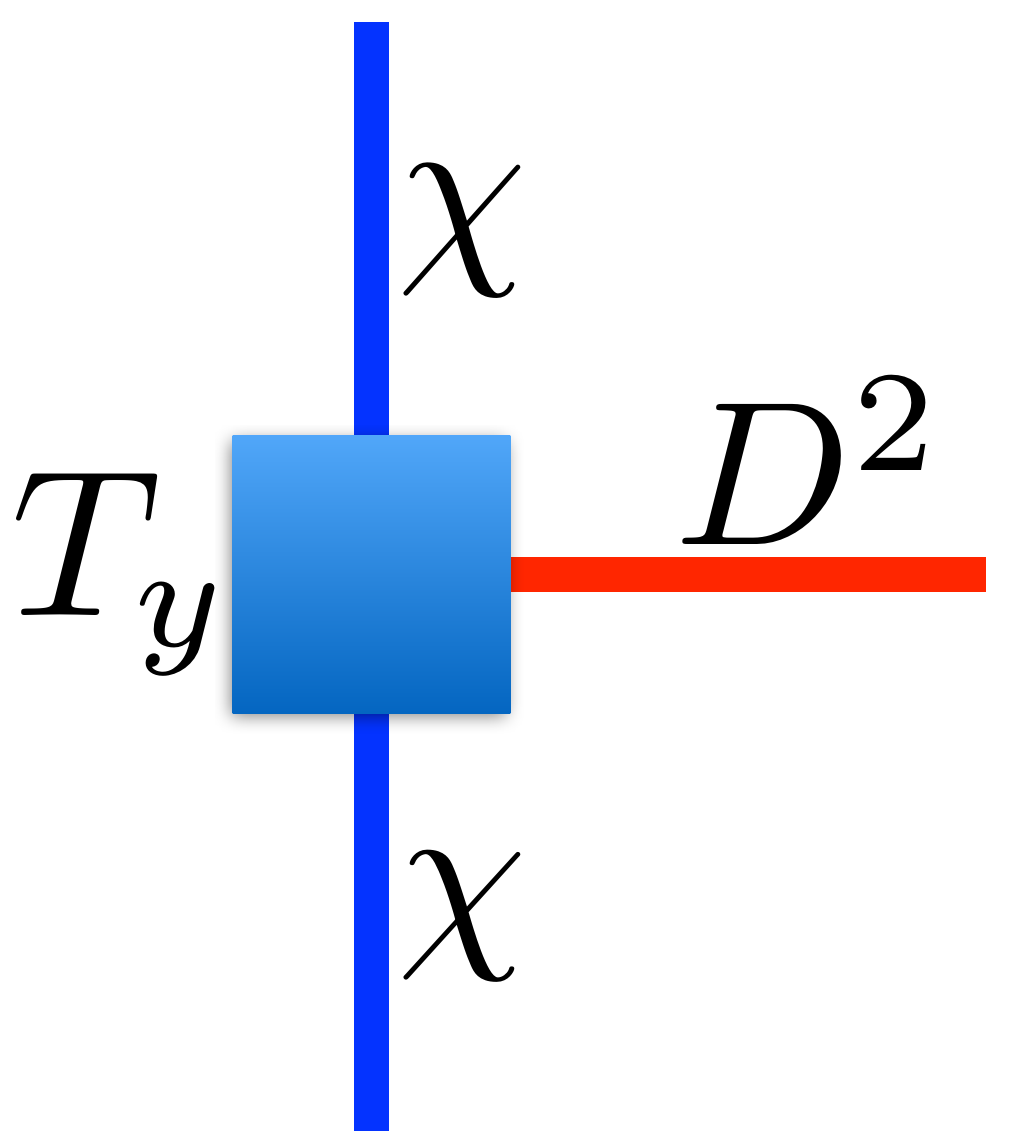}}\\
	\subfloat[]{\includegraphics[width=20.5mm, height=36.5mm]{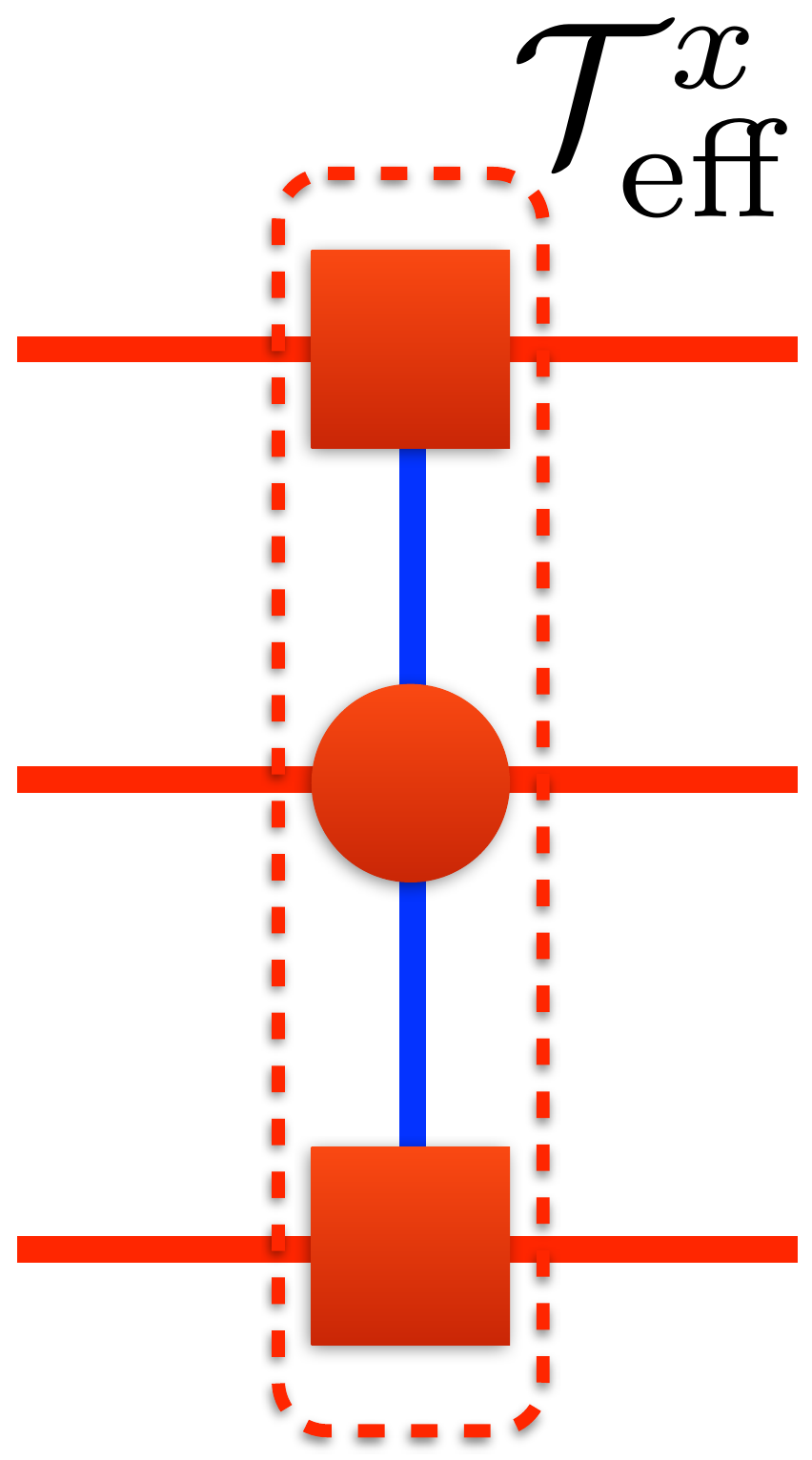}}\hspace{10mm}
	\subfloat[]{\includegraphics[width=41.5mm, height=20.5mm]{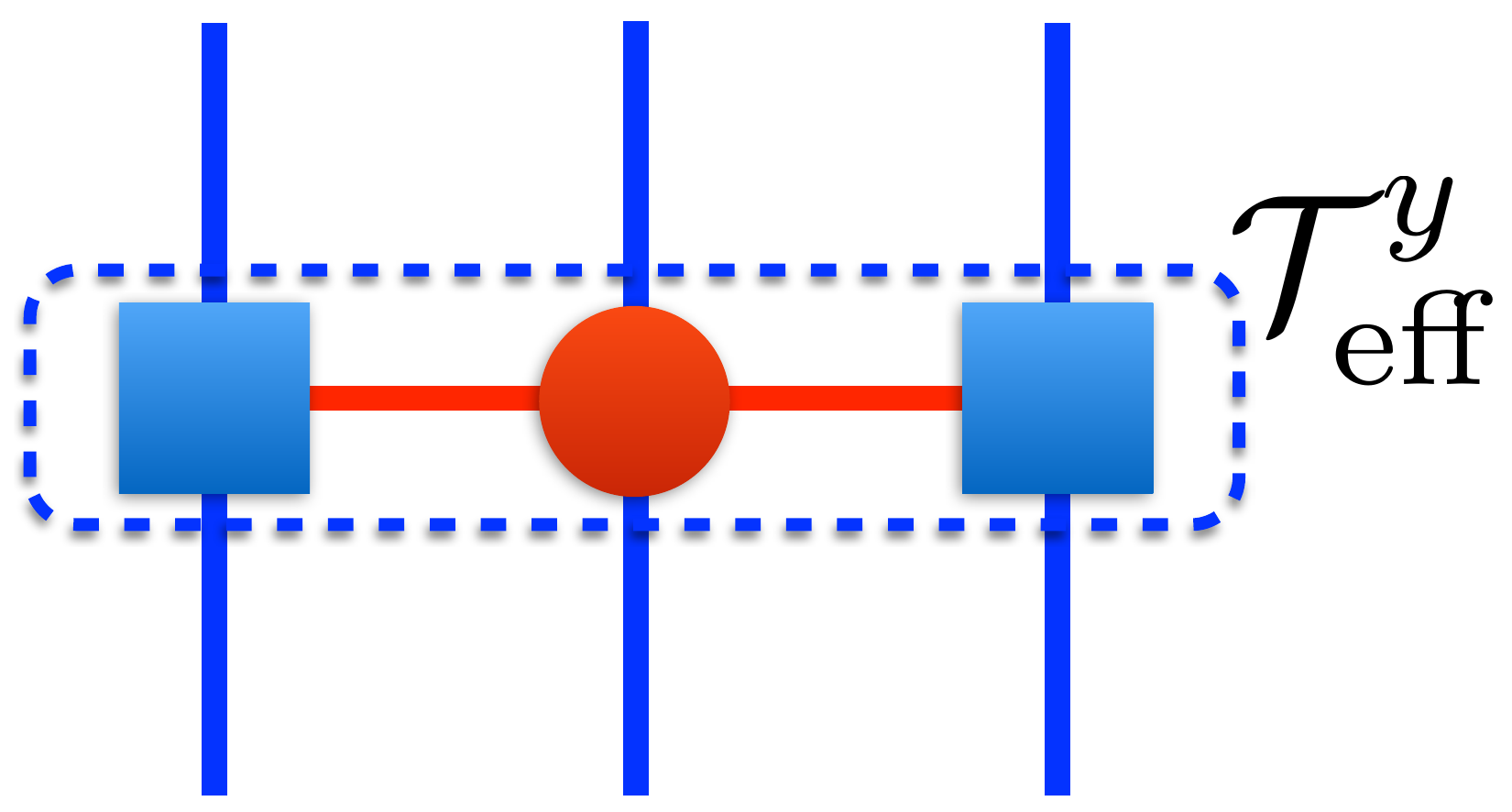}}
\caption{
[Color online] CTM RG for nematic PEPS with one-site unit cell. 
The $\pi/2$ lattice rotation symmetry breaking is indicated by red/blue line in horizontal/vertical direction. 
(a) Nematic PEPS tensor $\mathcal{A}$. (b) Double tensor $E$ obtained by tracing out physical indices $E=\sum_s\bar{\mathcal{A}}^s\mathcal{A}^s$, where $\bar{\mathcal{A}}^s$ is 
complex conjugate of $\mathcal{A}^s$. 
(c) CTM RG environment for $2\times 2$ cluster, constructed with corner matrix $C$, boundary tensor $T_{x, y}$ in horizontal/vertical direction, as shown separately in (d), (e), (f). 
The environment bond dimension is chosen to be $\chi=kD^2 (k\in \mathbb{N}_{+})$. 
Energy density is calculated by inserting either identity operator or the local hamiltonian operator in the red shaded $2\times 2$ cluster. 
(g), (h) is effective transfer matrix $\mathcal{T}^{x, y}_{\mathrm{eff}}$ in horizontal/vertical direction, constructed with $T_{x, y}$ and $E$ tensor. 
The maximal correlation length in horizontal/vertical direction can be obtained from largest two eigenvalues of $\mathcal{T}^{x, y}_{\mathrm{eff}}$.
}
\label{fig:CTMRG}
\end{figure}

{\it PEPS ansatz for the nematic phase --}
Both Lanczos ED and DMRG results suggest that, in the range $0< K_1 < 0.15$, there is no magnetic order. This motivates us to study this region with SU(2)-symmetric PEPS 
directly in the thermodynamic limit.  
To do so, we first apply a $\pi$-rotation along the $Y$ spin axis on one of the sublattices, and then 
approximate the GS wave function(s) in terms of a unique site tensor $\mathcal{A}_{\alpha\beta\gamma\delta}^{s}$,
where the greek indices label the states of the $D$-dimensional {\it virtual} space $\mathcal{V}$ attached to each site in the 
$z=4$ directions of the lattice,
and $s=0,\pm 1$ is the $S_z$ component of the physical spin-$1$. 
A (lattice) nematic state bears symmetry properties that greatly constrain the PEPS ansatz.
To construct such an ansatz, we use a classification of fully SU(2)-symmetric (i.e. singlet) PEPS proposed recently~\cite{Mambrini2016} 
in terms of the four virtual composite spins attached to each site and
the IRREP of the square lattice point group $C_{4v}$. In this setting, 
the virtual space $\mathcal{V}$ is a direct sum of SU(2) IRREPs.
Since the nematic state breaks only the 90-degree lattice rotation (but is invariant under axis reflection), 
the simplest adequate PEPS site tensors have the form 
\begin{equation}\label{eq:nematicPEPS}
\mathcal{A} = \mathcal{A}_1 + \mathcal{B}_1 = \sum_{a=1}^{N_A} \lambda_a\mathcal{A}_1^a + \sum_{b=1}^{N_B} \mu_b\mathcal{B}_1^b,
\end{equation} 
graphically shown in Fig.~\ref{fig:CTMRG}(a),
where the real elementary tensors $\mathcal{A}_1^a$ and $\mathcal{B}_1^b$ have the same set of virtual spins and transform according to 
the $A_1$ (i.e. fully symmetric or $s$-wave like) and $B_1$ 
(i.e. $d$-wave like) IRREP. $\lambda_{a}$ and $\mu_b$ are the corresponding coefficients, chosen to be real, and $N_{A,B}$ is the number of such elementary tensors in each class.
By reversing the overall sign of the coefficients $\mu_b$, the other degenerate GS can be obtained~\footnote{We note that, in some cases, Eq.~\eqref{eq:nematicPEPS} with a subset of the elementary tensors does not produce a nematic state but instead a fully symmetric state, although the tensor $\mathcal{A}$ breaks lattice rotation symmetry. But in general, the state represented by Eq.~\eqref{eq:nematicPEPS} is a nematic state.}.
These tensors have been tabulated in Ref.~\onlinecite{Mambrini2016} for $D\le 6$,
and their numbers for all virtual spaces $\mathcal{V}$ 
with good variational energy for the frustrated spin-1 Heisenberg model 
are listed in Table~\ref{TABLE:numbers}.
Following previous studies of the non-chiral or chiral frustrated spin-$\frac{1}{2}$ antiferromagnetic Heisenberg model (AFHM)~\cite{Poilblanc2017a,Poilblanc2017b}, we consider a general superposition of all tensors of each class, and the coefficients $\lambda_{a}, \mu_{b}$ are considered as
variational parameters. 
To further support the existence of a nematic SL phase, we find it is interesting to consider, for comparison, the 
sub-class of fully symmetric $\mathcal{A}_1$ PEPS constructed from the $\mathcal{A}_1^{a}$ tensors only,
i.e. setting $\mu_b=0$ in Eq.~(\ref{eq:nematicPEPS}). 
The sub-class of $\mathcal{B}_1$ PEPS constructed with only $\mathcal{B}_1^{b}$ tensors, although giving also fully symmetric states, does not provide good energies (see text below) and therefore is not considered here.
\begin{table}[hbt]
\renewcommand\arraystretch{1.5}
\caption{
Numbers of independent SU(2)-symmetric spin-$1$ tensors for the four different 
virtual spaces we consider, $D\le 6$. The third (fourth) column gives the number of $\mathcal{A}_1$ ($\mathcal{B}_1$) tensors and 
the last column is for the total number of tensors in the PEPS ansatz of the spin-$1$ nematic state. 
}
\label{TABLE:numbers}
\begin{center}
\begin{tabular}{ccccc}
    \hline 
    \hline
    \qquad\quad ${\mathcal{V}}$ \qquad\qquad & \qquad $D$ \qquad\qquad\qquad & $\mathcal{A}_1$ \qquad\qquad & $\mathcal{B}_1$ \qquad\quad & Total  \# \\
    \hline
    $\frac{1}{2}\oplus 0$ & 3 \qquad\qquad & 2 \qquad\qquad & 2 \qquad\qquad & 4 \\
    $\frac{1}{2}\oplus 0\oplus 0$ & 4 \qquad\qquad & 6 \qquad\qquad & 5 \qquad\qquad & 11 \\
    $\frac{1}{2}\oplus \frac{1}{2}\oplus 0$ & 5 \qquad\qquad & 12 \qquad\qquad & 13 \qquad\qquad & 25 \\
    $1\oplus\frac{1}{2}\oplus 0$ & 6 \qquad\qquad & 13 \qquad\qquad & 13 \qquad\qquad & 26 \\
    \hline
    \hline 
\end{tabular}
\end{center}
\end{table}

To explore the physical properties,  i.e. energy density and other observables, directly in the thermodynamic limit using infinite-PEPS (iPEPS), we apply 
corner transfer matrix (CTM) renormalization group (RG) techniques~\cite{Nishino1996,Nishino2001,Orus2009,Orus2012}, taking advantage of simplifications 
introduced by the use of point-group symmetric tensors~\cite{Poilblanc2017a}. 
At each RG step a truncation of the CTM is done
by keeping (at most) $\chi$ eigenvalues/singular values and preserving exactly the SU(2) multiplet structure. 
For a fully symmetric $\mathcal{A}_1$ PEPS (given by  all $\mu_b=0$ in Eq.~\eqref{eq:nematicPEPS}) the CTM is hermitian, and truncation can be done using ED, following Ref.~\onlinecite{Poilblanc2017a}.
For a genuine nematic state given by Eq.~\eqref{eq:nematicPEPS}, the CTM is not hermitian due to lattice rotation symmetry breaking, and we need to use a singular value decomposition instead.
The specific environment tensors for our nematic PEPS, obtained from CTM RG, are shown in Fig.~\ref{fig:CTMRG} (c-f), which include the corner matrix $C$ and the 
boundary tensors $T_{x, y}$ in the horizontal/vertical direction. For the fully symmetric $\mathcal{A}_1$ PEPS, the boundary tensors in both directions are actually the same~\cite{Poilblanc2017a}.
To optimize the coefficients in Eq.~\eqref{eq:nematicPEPS} 
\cite{Corboz2016, Vanderstraeten2016, Liu2017}, we perform a 
conjugate gradient (CG) method~\cite{Numerical2007}, starting with small $\chi$, and gradually increasing $\chi$ up to a maximum 
$\chi=\chi_{\mathrm{opt}}$.
In this process, we take $\chi=kD^2 (k\in \mathbb{N}_{+})$,
and, since the number of elementary tensors in each class is small (see Table~\ref{TABLE:numbers}), we are able to calculate the gradient using a simple finite difference method.
After optimization, we continue to perform CTM RG with several larger $\chi$ values 
to obtain physical observables 
and then, eventually, take the limit $\chi\rightarrow\infty$ (using a ``rigid" ansatz) by extrapolating the data.

\begin{figure}[htb]
\centering
\includegraphics[width=0.7\columnwidth,angle=-90]{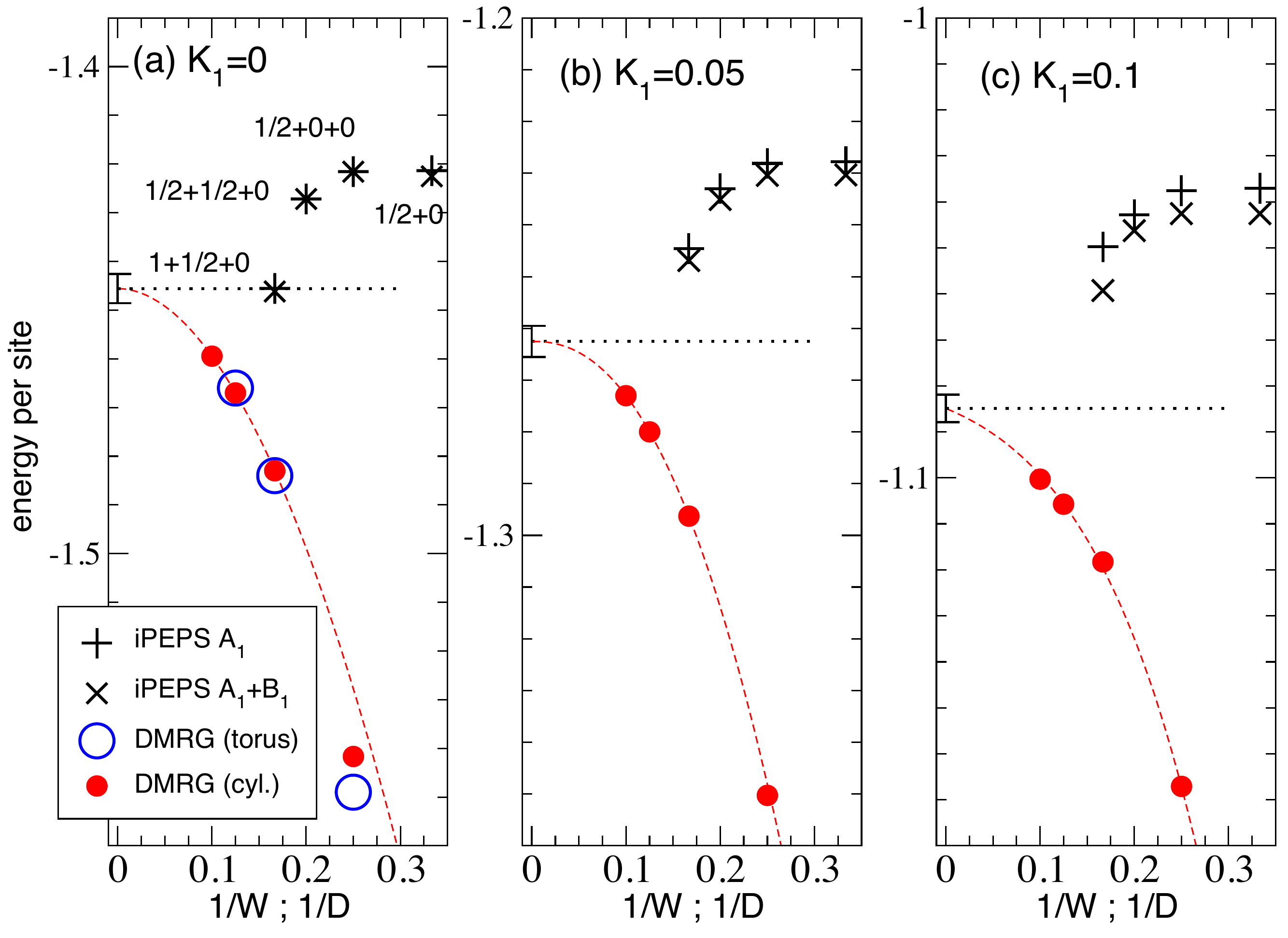}
\caption{
[Color online] iPEPS variational energies per site (extrapolated in the limit
$\chi\rightarrow\infty$) vs $1/D$. 
A comparison to DMRG data obtained on $2W\times W$ cylinders (see text) and plotted versus $1/W$ is shown.
In (a), data for $W\times W$ tori~\cite{Jiang2009} are also shown.
 }
\label{fig:energy_compare}
\end{figure}

Let us now first present results for the energy density at $K_1 = 0, 0.05, 0.1$. 
For all these cases, we shall use the full $\mathcal{A}_1 + \mathcal{B}_1$ ansatz 
of Eq.~\eqref{eq:nematicPEPS} as well as the restricted $\mathcal{A}_1$ 
symmetric ansatz.
We have explored all different classes of nematic state with $D\le 6$, and focus here on the virtual spaces 
$\mathcal{V} = \frac{1}{2}\oplus 0$, $\frac{1}{2}\oplus 0\oplus 0$, $\frac{1}{2}\oplus\frac{1}{2}\oplus 0$, $1\oplus\frac{1}{2}\oplus 0$, corresponding to $D=3$, $4$, $5$, $6$ respectively, with which the best variational energies have been found for each given value of $D$ (and $K_1=0$).
The energies for each class are plotted as a function of $D^2/\chi$ (see Appendix~\ref{app1}) and extrapolated 
to $\chi=\infty$.
The extrapolated PEPS energies are then compared to DMRG results in Fig.~\ref{fig:energy_compare} showing nice agreement, in particular for $K_1=0$. 
This indicates our ansatz provides good approximation for the true ground state in the thermodynamic limit, albeit with relatively small bond dimension $D$.

The fully symmetric $\mathcal{A}_1$ state with $D=3$ corresponds in fact to the so-called resonating Affleck-Kennedy-Lieb-Tasaki loop (RAL) state~\cite{Li2014}, 
and the larger $D$ case can be viewed as certain extensions of the RAL state. The nematic state is then obtained by mixing with $\mathcal{B}_1$ components and therefore breaking the lattice rotation symmetry.
It is interesting to note that, even with $D=3$, i.e. the RAL state, the energy is already fairly good, considering 
the fact that there is only one variational parameter to play with in the fully symmetric case. 
This is quite similar to the spin-1/2 case, where a one-parameter family of long-range Resonating Valence Bond (RVB) states gives also a good ansatz for the spin-1/2 $J_1-J_2$ AFHM~\cite{Poilblanc2017a}. 
Starting with $D=3, \mathcal{V} = \frac{1}{2}\oplus 0$, adding a spin-0 (increasing $D$ to $4$) barely changes the variational energy, as seen in Fig.~\ref{fig:energy_compare}. However, adding a virtual spin-$\frac{1}{2}$ ($D=5$) or a spin-1 ($D=6$) does significantly improve the energy. We also note that the energy gain by adding the $\mathcal{B}_1$ components (therefore breaking the $\pi/2$ lattice rotation symmetry) becomes larger with increasing $K_1$. 
This indicates that the nematic order becomes more prominent with increasing $K_1$ as discussed below.
\begin{figure}
\centering
	\subfloat[$D=3$]{
	\includegraphics[width=42mm, height=40mm]{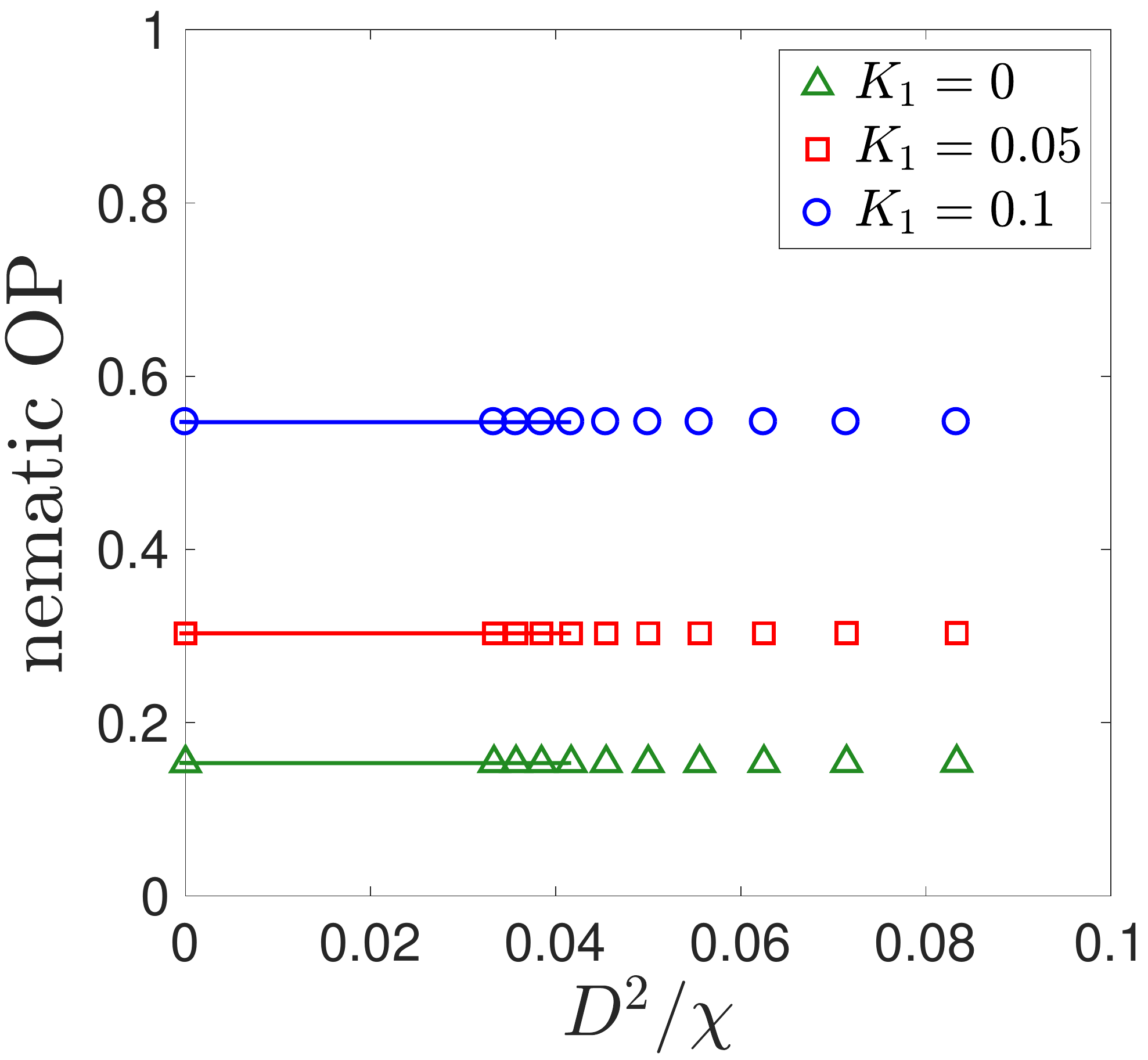}}
	\subfloat[$D=4$]{
	\includegraphics[width=42mm, height=40mm]{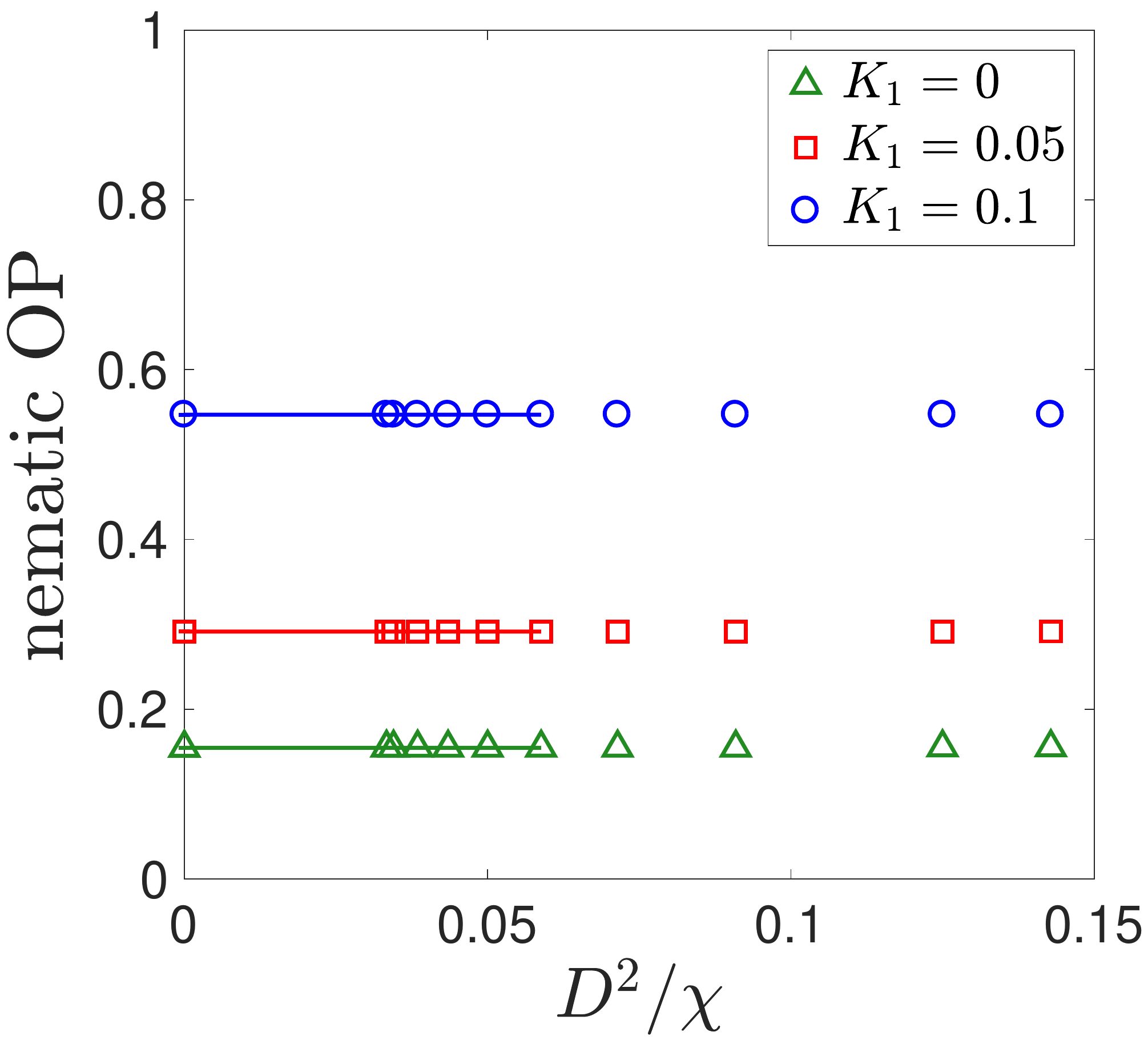}}\\
	\subfloat[$D=5$]{
	\includegraphics[width=42mm, height=40mm]{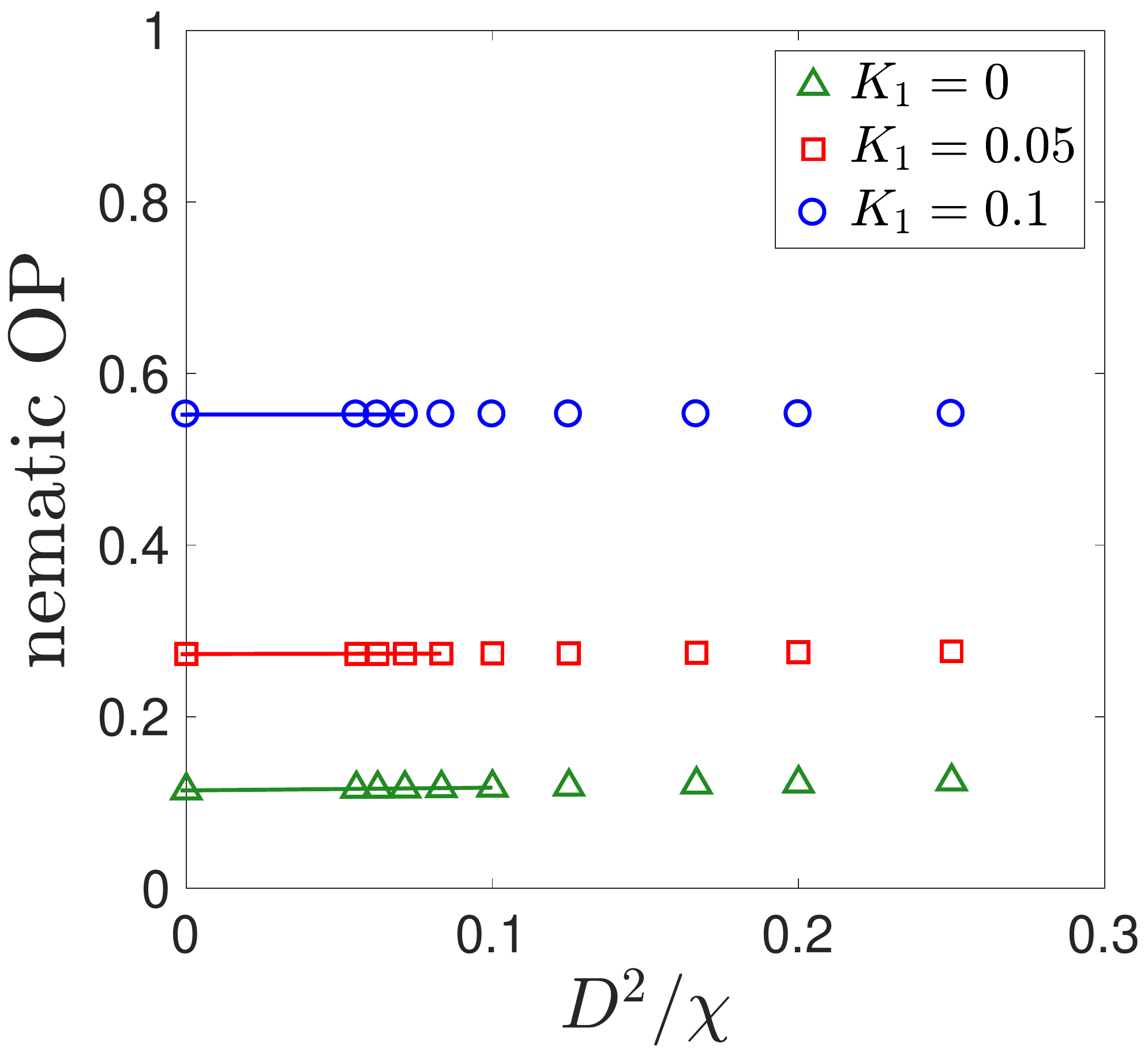}}
	\subfloat[$D=6$]{
	\includegraphics[width=42mm, height=40mm]{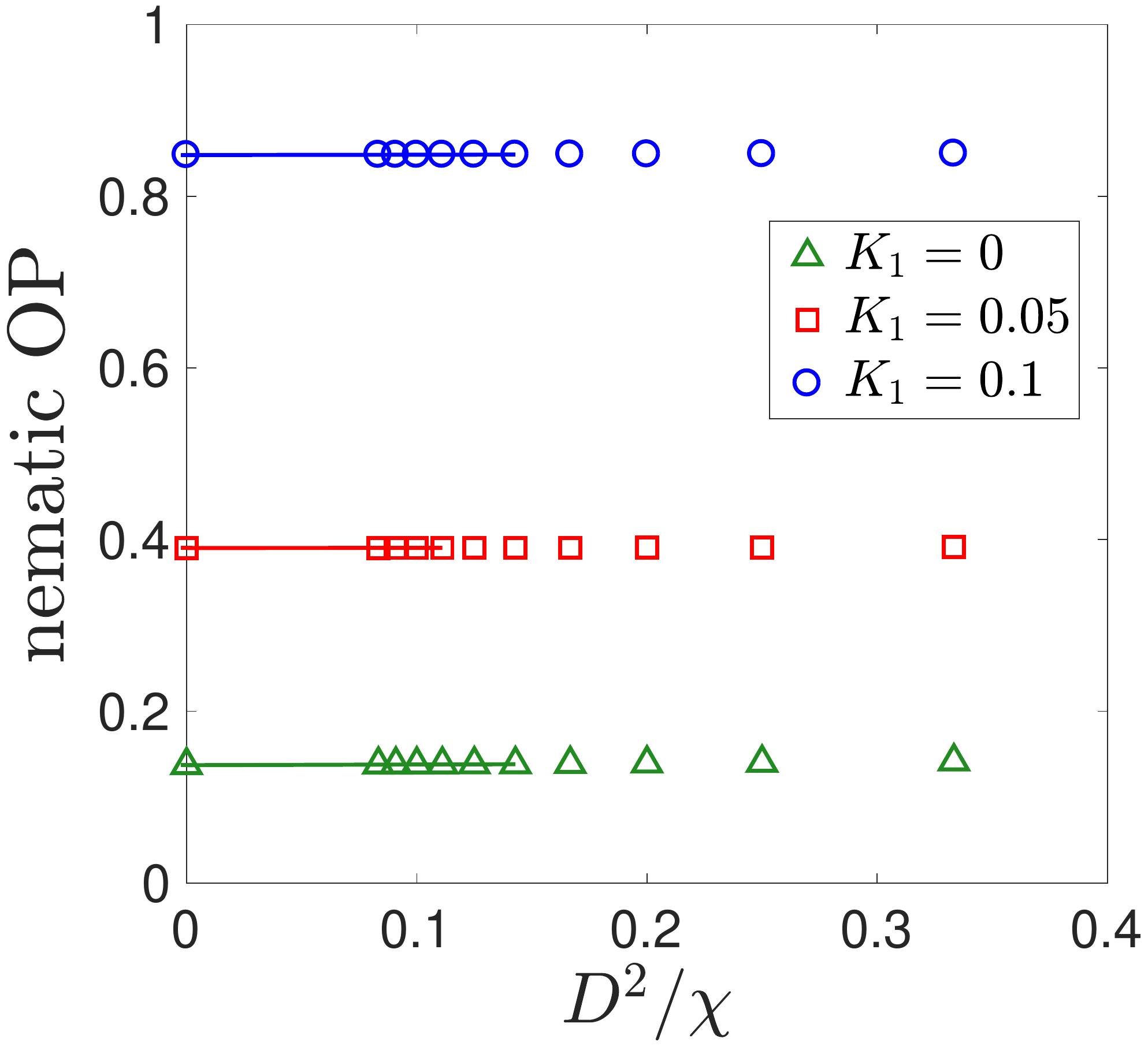}}\\
\caption{
[Color online] iPEPS nematic order parameter~\eqref{eq:nematic} for bond dimension $D$ from $3$ to $6$, plotted as a function of $D^2/\chi$. The extrapolated value ($\chi \rightarrow \infty$) is also shown, 
and the error bars are smaller than symbol size.
}
\label{fig:PEPS_nematicOP}
\end{figure}

Similar to energy density, the order parameter Eq.~\eqref{eq:nematic} associated with nematic PEPS Eq.~\eqref{eq:nematicPEPS}, 
can also be calculated using CTM RG environment tensors, with the same setting as in Fig.~\ref{fig:CTMRG} (c). The results are shown in Fig.~\ref{fig:PEPS_nematicOP}.
Unlike energy, which monotonically decreases with increasing $D$, the order parameter does not behave smoothly with $D$. 
Nevertheless, we can see that at $K_1=0$, the magnitude of the order almost shows no change with increasing $D$, 
while the changes are more significant for nonzero $K_1$, especially at $K_1=0.1$. More importantly, we find that, for each given bond dimension $D$, 
the order increases with $K_1$, showing full consistency with the increasing energy gain w.r.t.
the symmetric $\mathcal{A}_1$ PEPS.
This again shows that the $K_1$ term tends to stabilize the nematic phase.

We now explore further the physical properties of both the optimized nematic and symmetric PEPS wavefunctions, 
by looking at the maximal correlation length $\xi_{\mathrm{max}}$,
which can be obtained from the transfer matrix spectrum~\cite{Chen2017}, as shown in Fig.~\ref{fig:CTMRG} (g, h):
\begin{equation}
\xi_{\mathrm{max}} = -\frac{1}{\mathrm{ln}|\lambda_2/\lambda_1|},
\label{eq:maxCorrLength}
\end{equation}
where $\lambda_{1, 2}$ are the largest two eigenvalues of the effective transfer matrix $\mathcal{T}_{\mathrm{eff}}$, 
which are symmetric matrices due to reflection symmetry. Note that, $\lambda_1$ is in general non-degenerate, and $|\lambda_2|$ is strictly smaller than $|\lambda_1|$, 
consistent with the absence of long-range order in each state.
Since the horizontal and vertical directions are non-equivalent in a nematic state, we naturally have two different maximal correlation length $\xi^{x}_{\mathrm{max}}$ and
$\xi^{y}_{\mathrm{max}}$.

\begin{figure*}[htbp]
\centering
	\subfloat[$K_1=0$]{
	\includegraphics[width=43.5mm, height=43mm]{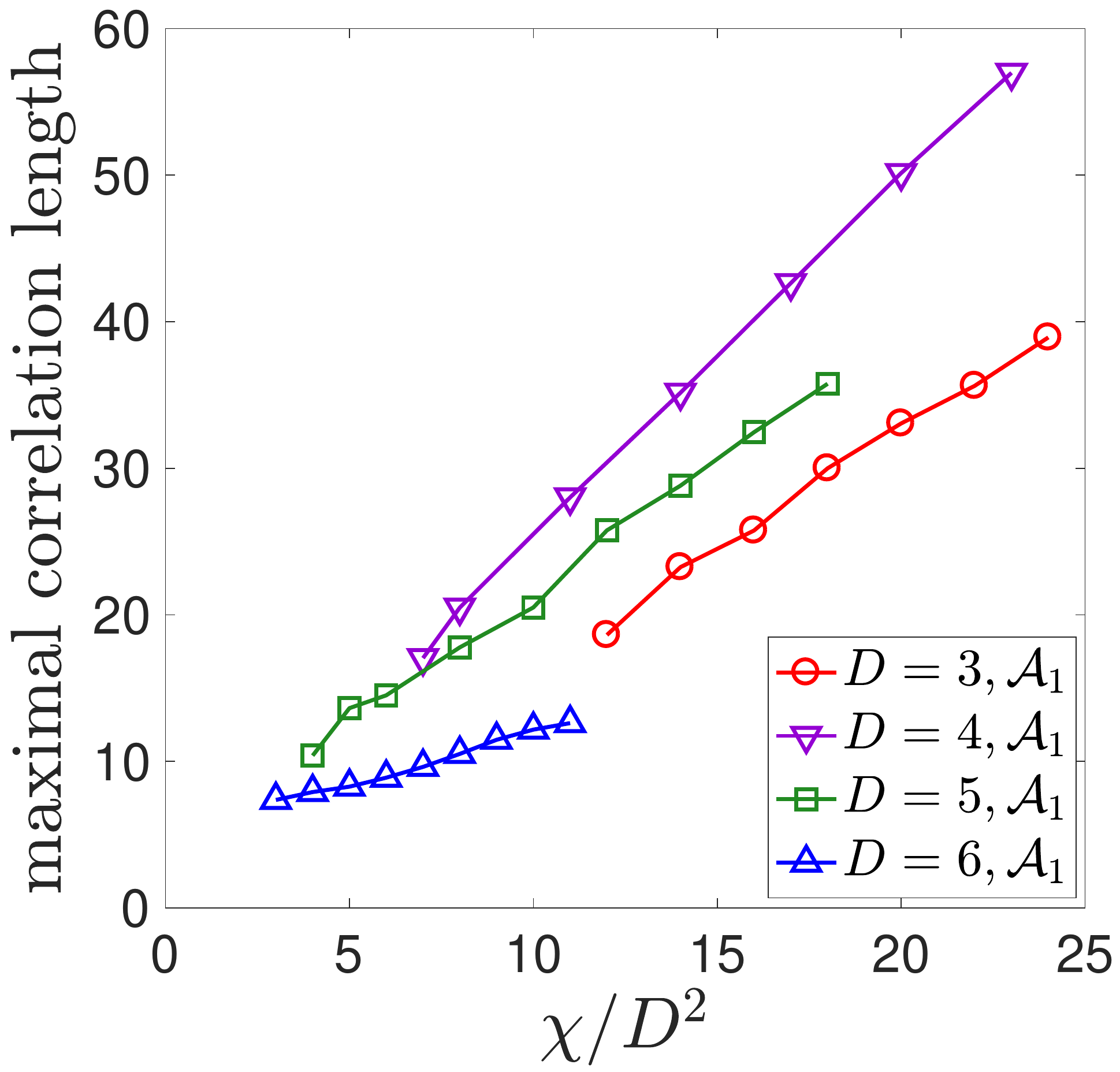}}
	\subfloat[$K_1=0$]{
	\includegraphics[width=43.5mm, height=43mm]{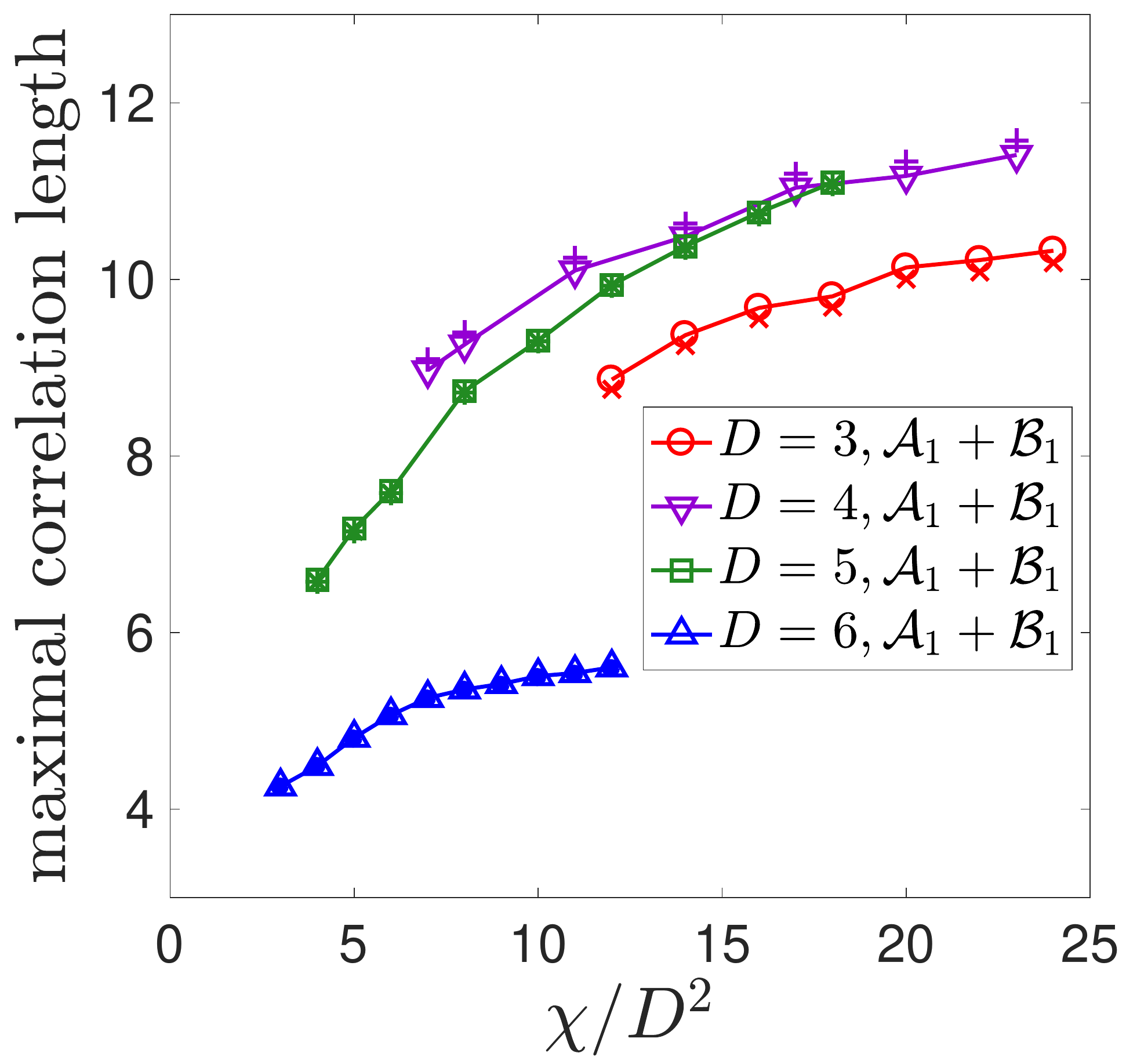}}
	\subfloat[$K_1=0.05$]{
	\includegraphics[width=43.5mm, height=43mm]{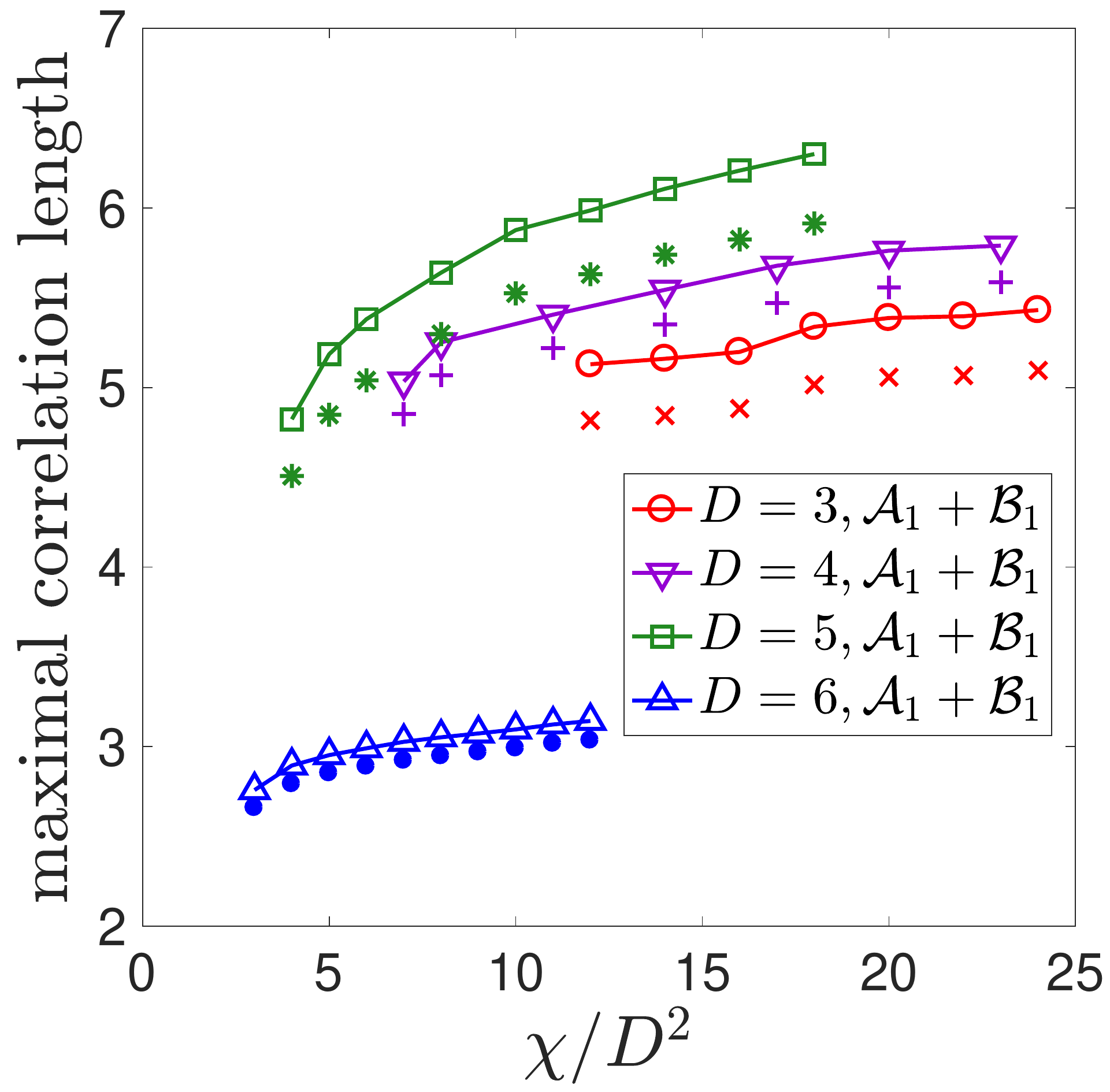}}
	\subfloat[$K_1=0.1$]{
	\includegraphics[width=43.5mm, height=43mm]{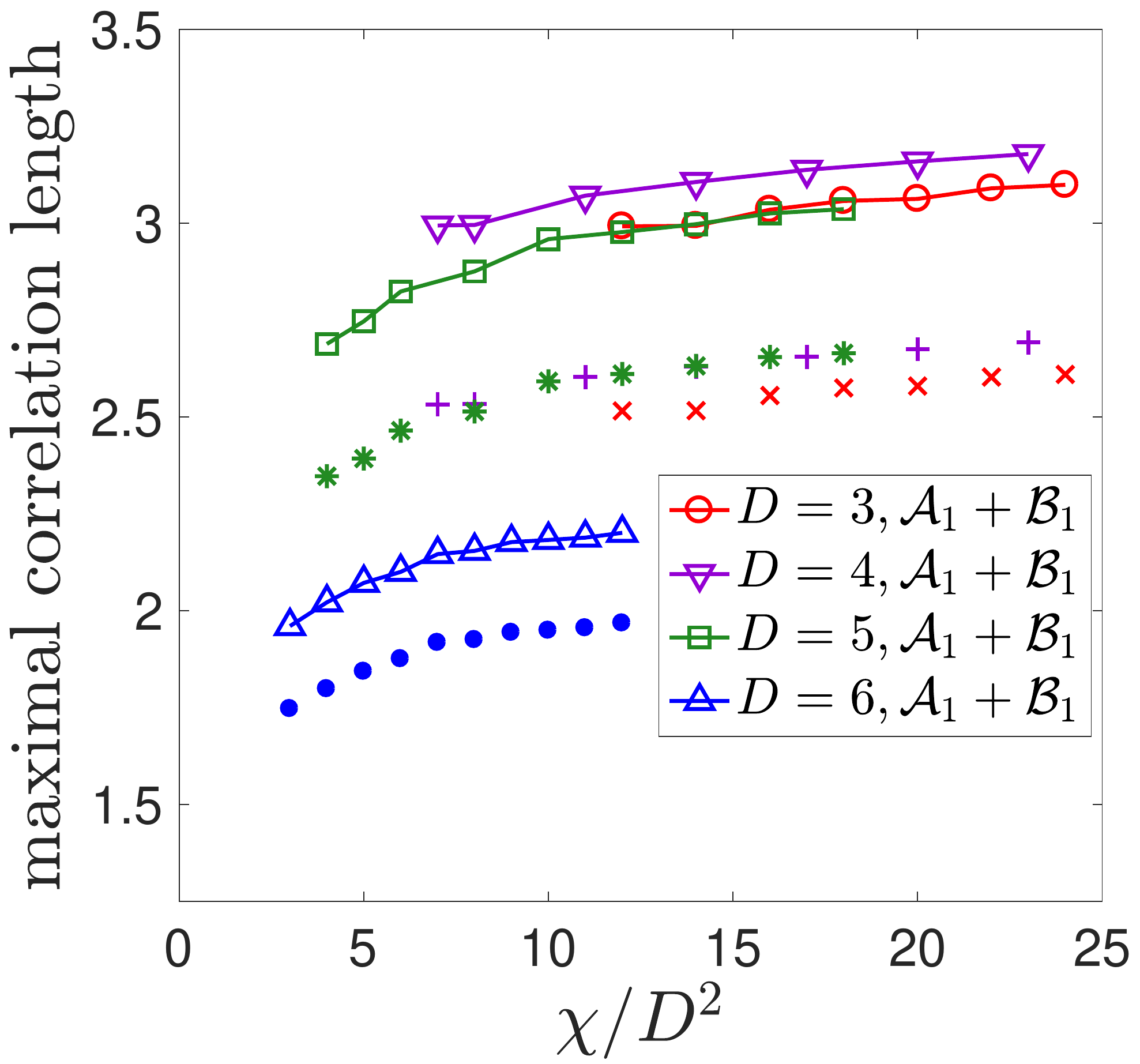}}
\caption{
[Color online] iPEPS maximal correlation length $\xi_{\mathrm{max}}$ in the symmetric PEPS (a) and in the nematic PEPS (b,c,d) versus $\chi/D^2$, for different values of $K_1$. The data for different values of $D$ ranging from $3$ to $6$ are shown with different symbols according to the legends.
(a,b) $K_1=0$;  (c) $K_1=0.05$; (d) $K_1=0.1$. In (b-d), open / filled symbols with same color correspond to the weak / strong bond directions of nematic state. 
}
\label{fig:PEPS_maxCorrLength}
\end{figure*}

The optimized fully symmetric $\mathcal{A}_1$ states are all critical with diverging correlation length, as revealed in Fig.~\ref{fig:PEPS_maxCorrLength}(a) for $K_1=0$ and in Appendix~\ref{app2} for other values of $K_1$, showing a linear increase of the correlation length $\xi_{\mathrm{max}}$ with environment bond dimension $\chi$, with no sign of saturation. The criticality of $D=3$ $\mathcal{A}_1$ state, i.e. RAL state, has been studied in Ref.~\cite{Li2014}, where it was found that, on the square lattice, the RAL state has exponentially decaying spin-spin correlation function but algebraically decaying dimer-dimer correlations. This is similar to the property of the NN RVB state~\cite{Chen2017}, the criticality of both states being related to the exact U(1) gauge symmetry they possess. For the RAL state, this U(1) gauge symmetry is connected to the fact that there are always exactly two virtual dimers attached to every site. 
The criticality of the $D=4$ $\mathcal{A}_1$ state follows from the one of the RAL state, since only a spin-$0$ is added to the $D=3$ virtual space, and the property of having exactly two virtual dimers on every site is preserved. Therefore, it is also related to the same U(1) gauge symmetry.
For the $D=5$ ($D=6$) $\mathcal{A}_1$ states with virtual space $\mathcal{V}=\frac{1}{2}\oplus\frac{1}{2}\oplus 0$ ($1\oplus\frac{1}{2}\oplus 0$), only the {\it parity} of the number of virtual dimers is a good quantum number, leading to a smaller $\mathbb{Z}_2$ gauge symmetry, from which one would (naively) expect a finite correlation length. However, the linear increase of the correlation length 
$\xi_{\mathrm{max}}$ with $\chi$ strongly suggests 
that these PEPS are also critical. These features are quite similar to the ones of the long-range RVB state~\cite{Chen2017}, which bears both a discrete $\mathbb{Z}_2$ gauge symmetry and critical (dimer) correlations.
It is likely that the $D=5,6$ $\mathcal{A}_1$ states also have {\it emergent} U(1) gauge symmetry, responsible for their critical nature.

In deep contrast, the nematic states (Eq.~\eqref{eq:nematicPEPS}) all have finite correlation lengths, which clearly saturate with increasing $\chi$, as shown in Fig.~\ref{fig:PEPS_maxCorrLength}(b, c, d) for $K_1=0, 0.05, 0.1$, respectively. This behavior is consistent with symmetry breaking, 
which lowers the energy by inducing a nonzero order parameter and would develop a gap (in the spectrum of any local parent Hamiltonian). 
The magnitude of the energy gain is proportional to the size of the order parameter and,
approximately, to the size of the gap, which is itself inversely proportional to the correlation length. 
Thus, the decrease of the (saturated) correlation length with increasing $K_1$ (and the increase of the difference between $\xi^{x}_{\mathrm{max}}$ and $\xi^{y}_{\mathrm{max}}$) is consistent with (i) the increase of the energy gain w.r.t. the symmetric (critical) PEPS and (ii) the increase of the nematic OP, as we can see in Fig.~\ref{fig:energy_compare} and Fig.~\ref{fig:PEPS_nematicOP} respectively.
This also indicates that $K_1=0$ is close to a critical point (even though the actual transition with the N\'eel state may be weakly first order), and that the nematic phase becomes more stable with increasing $K_1$, before the stripe magnetic order sets in abruptly.
It is interesting to observe that, in contrast, $\xi_{\mathrm{max}}$ of the symmetric state diverges faster (see Appendix~\ref{app2}) with increasing $K_1$. 
It is also worth mentioning that, even in the presence of an exact U(1) gauge symmetry, the nematic state has finite correlation length, as we have checked for linear combinations of
$D=3$ $\mathcal{A}_1$ and $\mathcal{B}_1$ tensors~\footnote{The $D=3$ $\mathcal{A}_1$ class contains two tensors, both of which have two virtual spin-$1/2$ on every site configuration. The $D=3$ $\mathcal{B}_1$ class contains two tensors, denoted as $\mathcal{B}_1^{(1)}$ with two virtual spin-$1/2$s on every site configuration and $\mathcal{B}_1^{(2)}$ with four virtual spin-$1/2$s on every site configuration. The PEPS constructed by mixing $\mathcal{A}_1$ and $\mathcal{B}_1^{(1)}$ tensors has U(1) gauge symmetry, but also a finite correlation length.}.

Finally, we note that, although the integer nature of the physical spin together with the spin-SU(2) symmetry formally leads to a $\mathbb{Z}_2$ gauge symmetry of the tensor (with the nontrivial group element generated by $2\pi$ SU(2) spin rotation~\cite{Jiang2015}), we have checked explicitly using the tensor renormalization group method~\cite{He2014, Mei2017, Chen2017} that the modular matrices of our optimized (short-ranged) nematic states are actually trivial, implying the absence of topological order in this system.

{\it Summary and outlook.} 
We have studied a nematic spin liquid phase in the frustrated spin-1 model with three complementary numerical methods: ED, DMRG and iPEPS.
Both ED and DMRG results suggest there exists a non-magnetic phase which breaks $\pi/2$ lattice rotation symmetry 
in the parameter region $0< K_1<0.15$, for fixed $J_1=1, J_2=0.54$. However, since ED can only deal with small system sizes, and the cylinder geometry used in DMRG breaks lattice rotation symmetry, it is important to use a complementary method which (i) works directly in the thermodynamic limit and (ii) enables to compare symmetric and symmetry-broken states. 
Hence, to get new insight on this plausible nematic phase, 
we have applied the iPEPS method to study the relevant parameter region directly in the thermodynamic limit. 

Based on a previous classification of SU(2) spin rotation symmetric tensors~\cite{Mambrini2016}, we systematically construct variational fully symmetric PEPS and nematic PEPS for the frustrated spin-1 Heisenberg model, which are all singlet states, targeting the correct non-magnetic phase (without 
necessarily resorting to certain scaling with bond dimension $D$). 
The nematic PEPS is obtained by superposing two different classes of tensors, i.e., $\mathcal{A}_1$ and $\mathcal{B}_1$, while the fully symmetric PEPS is constructed only with $\mathcal{A}_1$ tensors.
Through a comparison study with fully symmetric $\mathcal{A}_1$ PEPS and nematic $\mathcal{A}_1+\mathcal{B}_1$ PEPS, we unambiguously demonstrate the existence of the nematic phase. The PEPS variational energies agree very well with DMRG, therefore validating our ansatz. Through a detailed analysis of the nematic order parameter and the (maximal) correlation length of both symmetric and nematic PEPS, we find that the positive NN biquadratic term stabilizes the nematic phase, in agreement with ED and DMRG.
It is interesting to note that the same nematic phase was also found with only NN bilinear and biquadratic terms on square lattice~\cite{Niesen2017}.

In summary, we have used a simple classification of SU(2) invariant PEPS to construct 
a generic family of  well controlled ans\"atze of nematic spin liquids. 
The physical relevance of such states for a simple frustrated spin-1 Heisenberg model is established by direct comparisons to unbiased 
Lanczos ED and DMRG calculations.
Such an approach could easily be extended to investigate many other types of lattice symmetry-breaking non-magnetic phases in frustrated quantum spin models, and is left for future investigations.

Lastly, we point out that a possibly continuous transition between the $(\pi, \pi)$ N\'eel state and the nematic state 
can be described as a deconfined critical point of a non-compact CP$^1$
model~\cite{Senthil2004} which is conjectured to possess an emergent $O(4)$ symmetry~\cite{Metlitski2017,Wang2017_2}. 
However, our numerics cannot fully decide whether the transition is indeed continuous or 
weakly first order. Nevertheless, the symmetric critical $\mathcal{A}_1$ PEPS may provide a good representation of the critical point. Such an investigation is left for a future study. 

{\it Note added ---} Upon finishing this work, we became aware of related work~\cite{Haghshenas2018}.

\begin{acknowledgements}
This project is supported by the TNSTRONG
ANR grant (French Research Council).  This work was granted access to the HPC resources of CALMIP and GENCI supercomputing centers under the allocation 2017-P1231 and A0030500225 respectively. We acknowledge inspiring conversations with S.-S. Gong, P.~Pujol, S.~Sachdev and D. N. Sheng. JYC thanks H.-H. Tu for interesting discussion. We thank Hong-Chen Jiang for sending us the torus DMRG data for comparison, and 
thank Centro de Ciencias de Benasque Pedro Pascual for hospitality, where this work was finalized.

\end{acknowledgements}

\bibliography{bibliography}

\appendix
\renewcommand\thefigure{\thesection.\arabic{figure}}
\setcounter{figure}{0}

\section{Additional data from DMRG simulations}
\label{app:dmrg}

\begin{figure}
\begin{center}
\includegraphics[width=\columnwidth,angle=0]{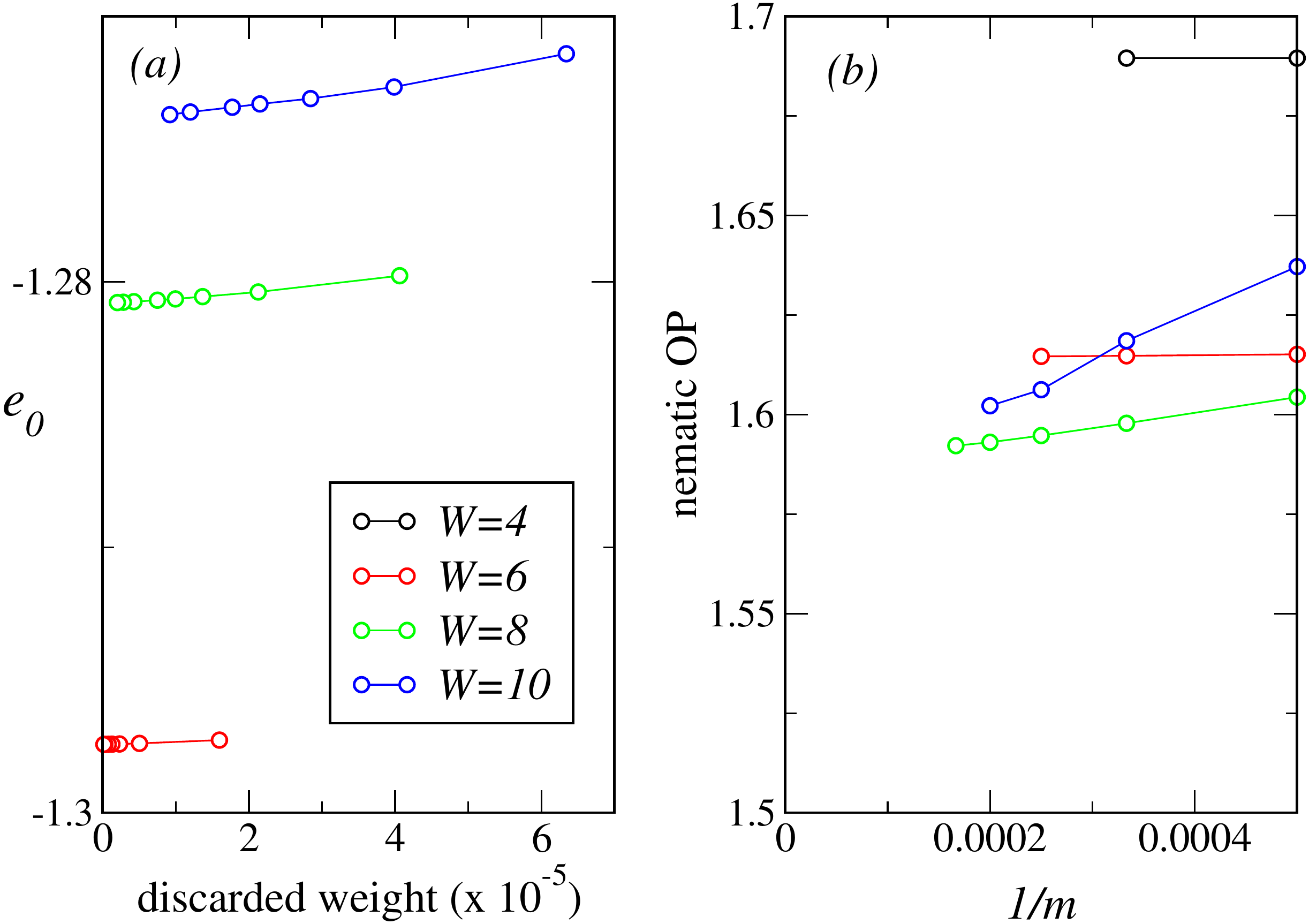}
\caption{
[Color online] Extrapolations performed for DMRG simulations on cylinder $2W\times W$ at fixed $K_1=0.05$. (a) The ground-state energy per site $e_0$ is extrapolated w.r.t. the discarded weight. (b) The nematic order parameter measured in the bulk of the system is extrapolated w.r.t. $1/m$.
 }
\label{fig:dmrg_extrap}
\end{center}
\end{figure}

\subsection{Convergence of energy and nematic OP}
As is usually done in DMRG, for each given $L\times W$ cylinder one can attempt to extrapolate ground-state total energy w.r.t. the discarded weight, see Fig.~\ref{fig:dmrg_extrap}(a). In order to reduce the effects of open boundary conditions along the long direction ($L$), it is also suitable to perform simulations on different cylinders of length $L_1$ and $L_2>L_1$ and obtain an estimate of the ground-state energy density using:
$$ e_0(W) = \frac{E_0(L_2,W)-E_0(L_1,W)}{(L_2-L_1)W}$$
Note that when performing such extrapolations, the estimate is not variational anymore. We have used this procedure to get the best estimate of $e_0$ to be compared to iPEPS values. 

Regarding the nematic order parameter~(\ref{eq:nematic}), for a fixed cylinder, we have chosen to extrapolate its value w.r.t. $1/m$ where $m$ is the number of states kept in the DMRG (up to $6,000$ in our largest simulations), see Fig.~\ref{fig:dmrg_extrap}(b). Moreover, for a fixed width $W$, we have checked that its value is rather independent on $L$ so that even shorter cylinders (i.e. $L<2W$) could be used (data not shown). 

\begin{figure*}
\centering
	\subfloat[]{
	\includegraphics[width=60mm]{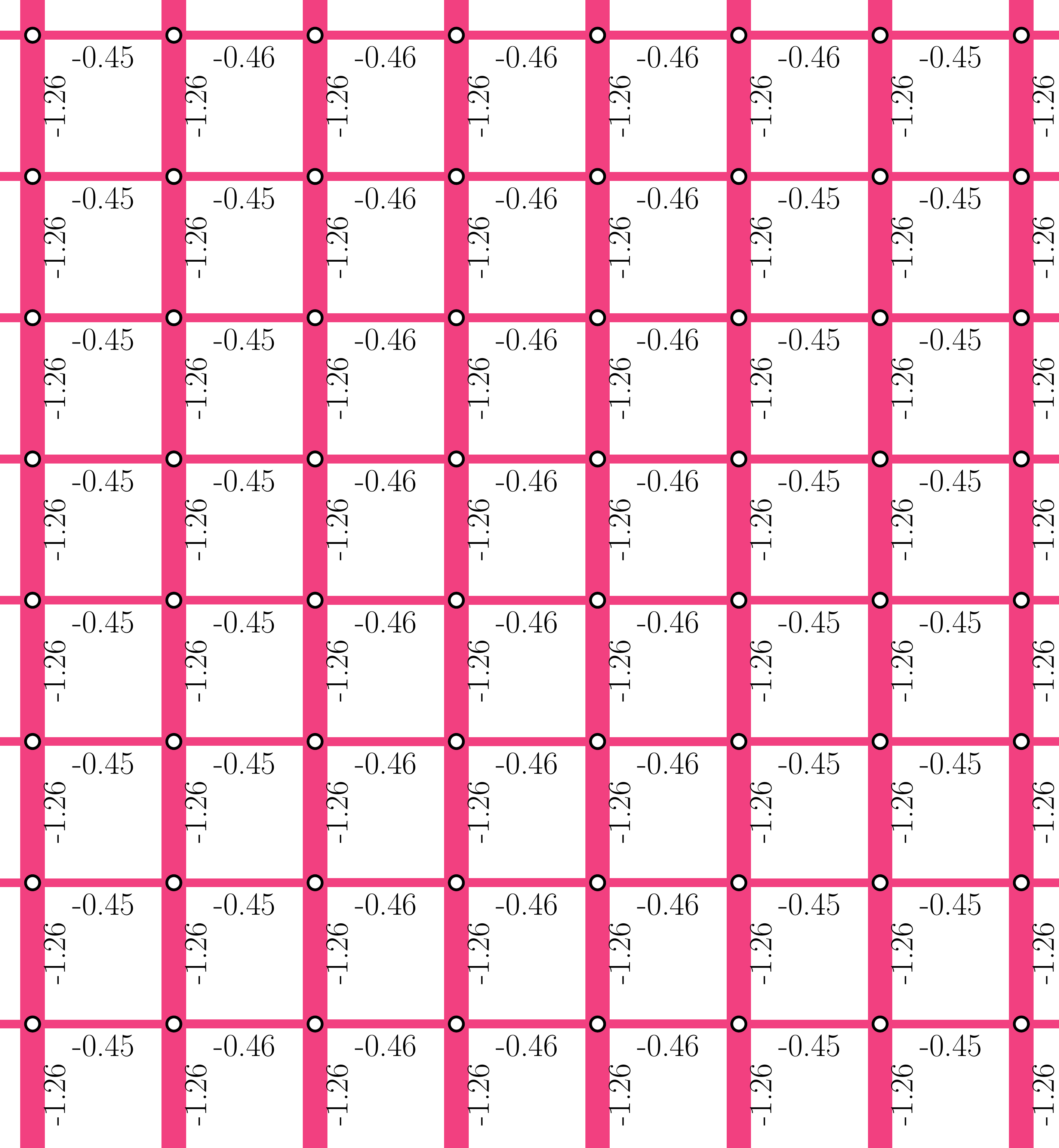}}
	\subfloat[]{
	\includegraphics[width=60mm]{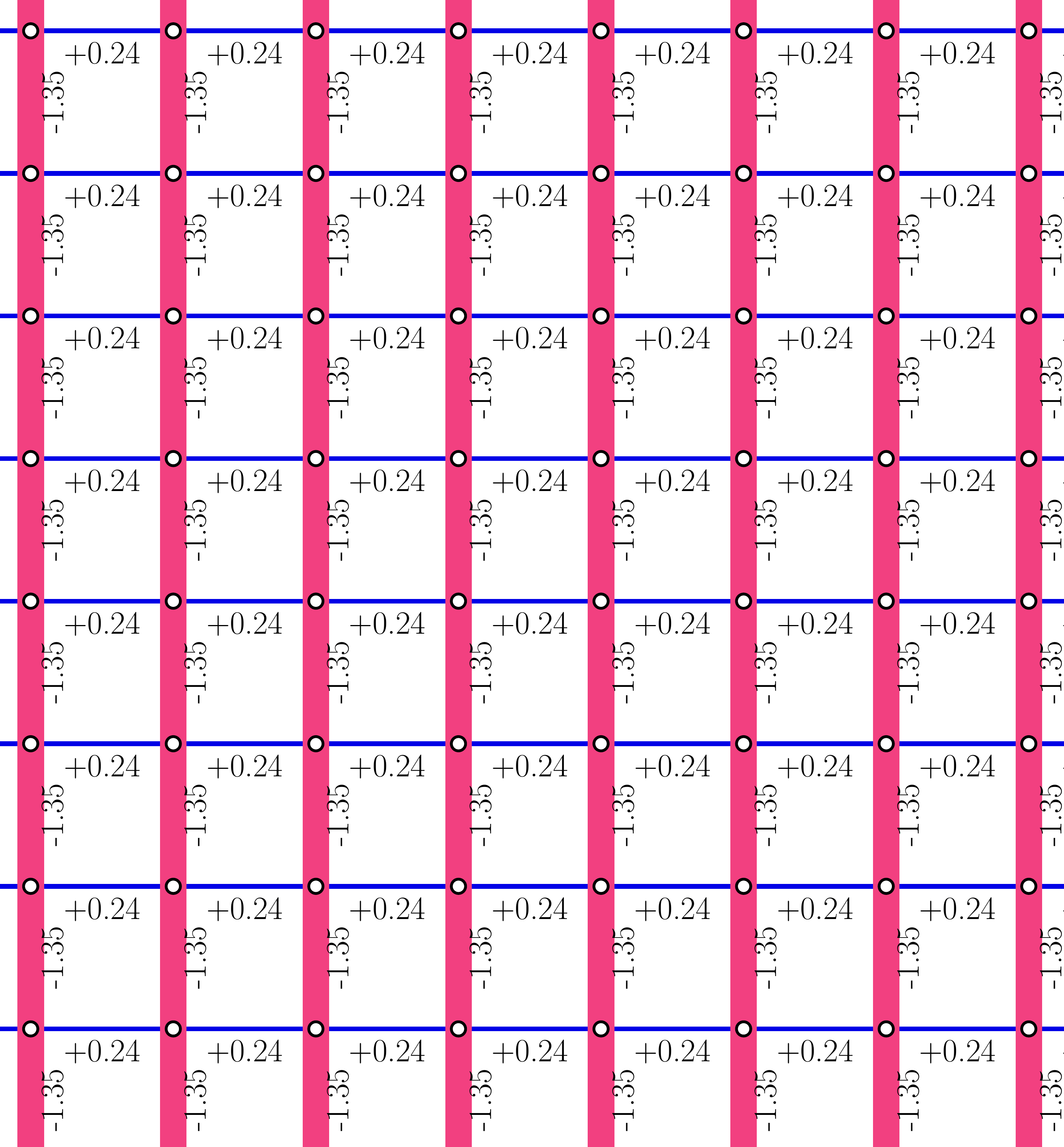}}
	\subfloat[]{
	\includegraphics[width=60mm]{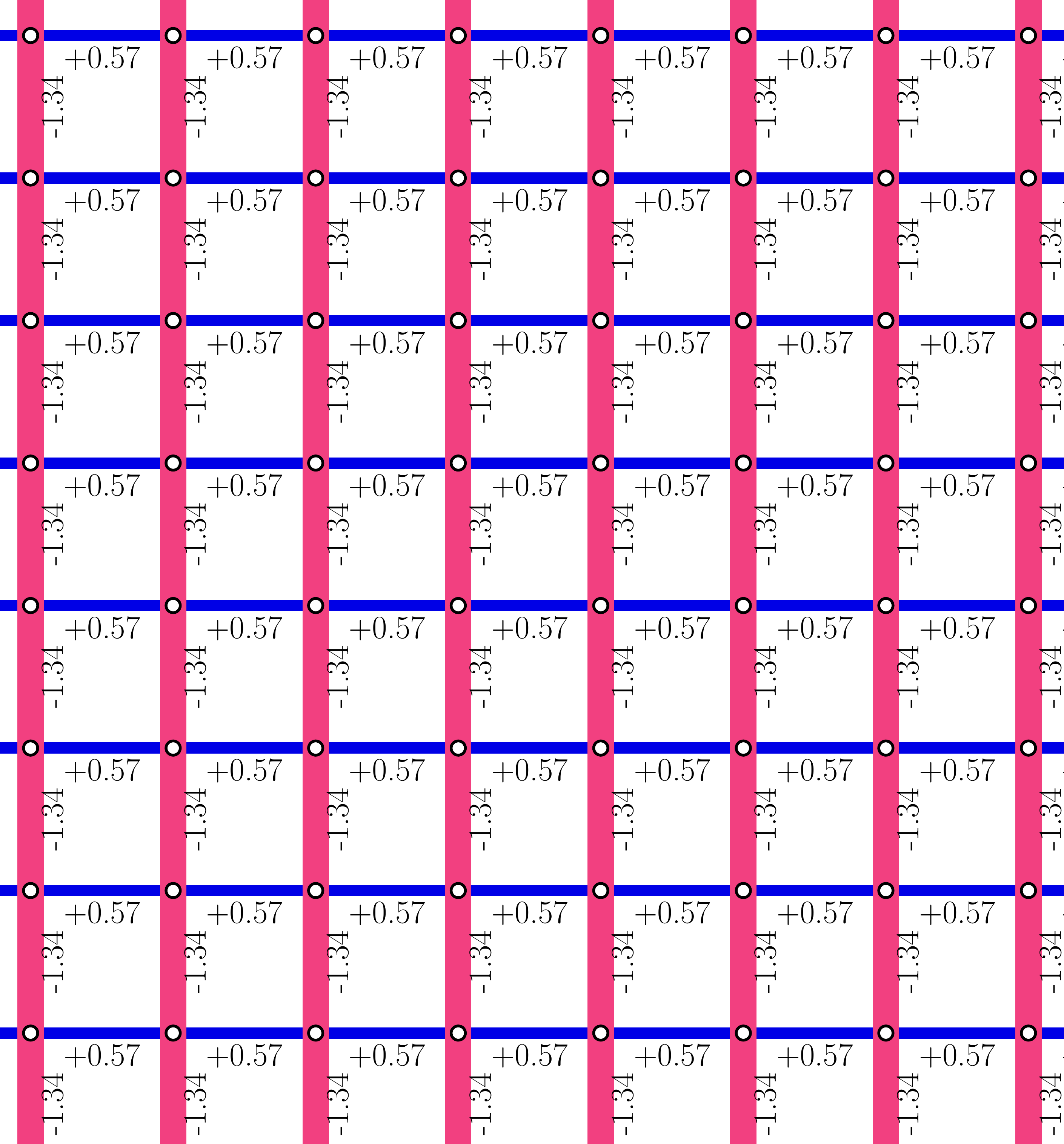}}
\caption{
[Color online] Bond energies obtained by DMRG on a $16\times 8$ cylinder keeping up to $m=5,000$ states.  
(a) $K_1 = 0$. (b) $K_1 = 0.05$. (c) $K_1 = 0.1$.
}
\label{fig:dmrg_bonds}
\end{figure*}

\subsection{Bond energies on a finite cylinder}

By simply measuring the bilinear bond energies ($\langle  {\bf S}_i\cdot{\bf S}_j\rangle$ for neighboring $(i,j)$ sites), we can already visualize a qualitative change when $K_1\gtrsim 0$, see Fig.~\ref{fig:dmrg_bonds}. In this range of parameters, the horizontal bonds become slightly ferromagnetic and quite different from the vertical ones, hence suggesting a possible nematicity. Nevertheless, a cautious analysis is needed since these data were obtained on cylinders (with an aspect ratio of 2) so that horizontal and vertical bonds are not equivalent, see a detailed analysis below and in the main text.

\subsection{Finite-size scaling of the order parameters}

\begin{figure}[htbp]
\begin{center}
\includegraphics[width=\columnwidth,angle=0]{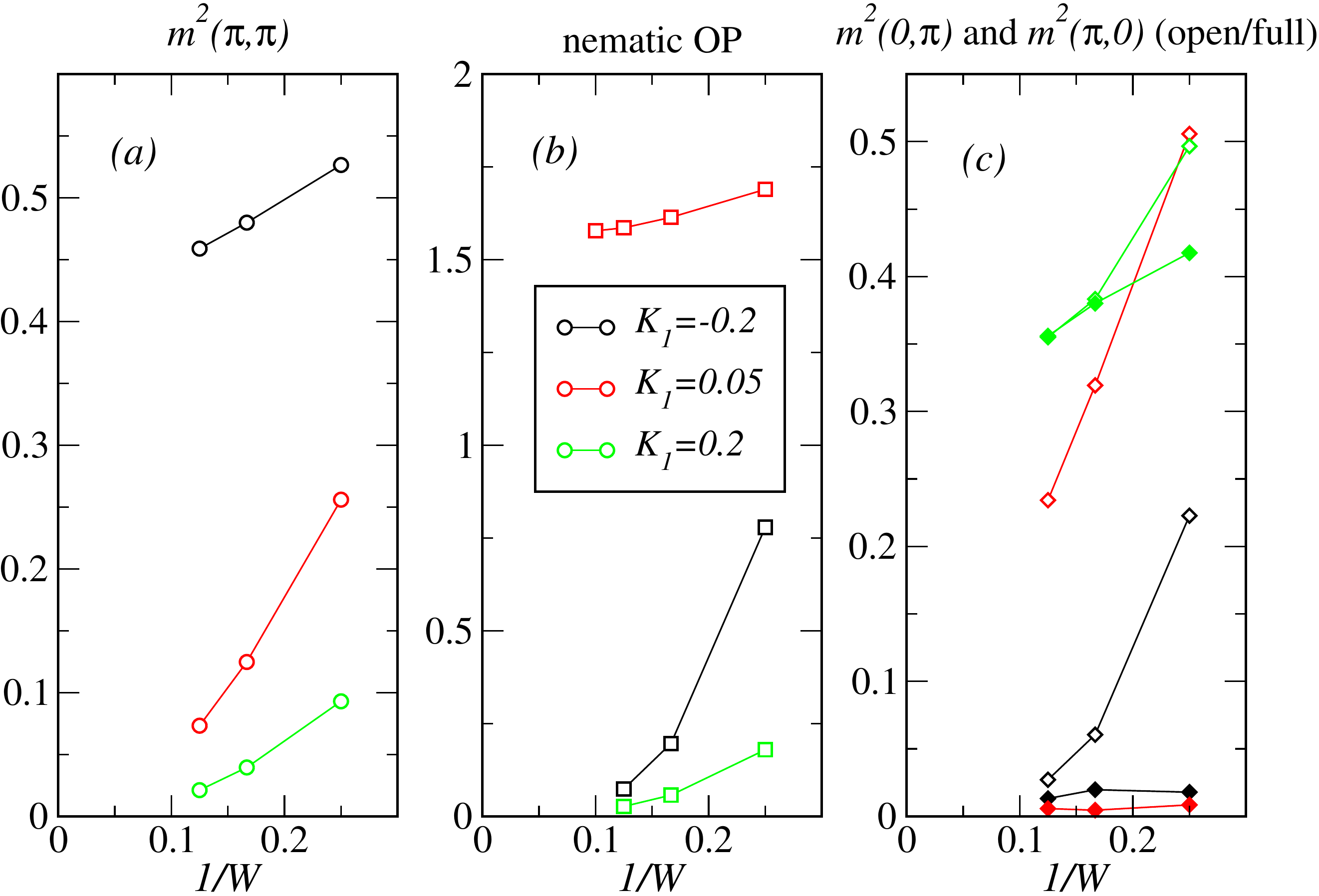}
\caption{
[Color online] Finite-size scaling of DMRG data obtained on cylinders $2W\times W$ vs $1/W$ for three typical values of $K_1=-0.2$, $0.05$ and $0.2$ corresponding respectively to the N\'eel, nematic and stripe phase.
 }
\label{fig:dmrg_scaling}
\end{center}
\end{figure}

As defined in the main text, we can compute several order parameters for the putative phases, but we do need to perform some finite-size scaling analysis. We have measured spin structure factor and nematic order parameter in the bulk (i.e. in the center) of each $2W\times W$ cylinder in order to reduce finite-size effects. In Fig.~\ref{fig:dmrg_scaling}, we provide some scaling analysis vs $1/W$. In particular, when $K_1=0.2$, the structure factor at  wavevector $(0,\pi)$ is much stronger than $(\pi,0)$ due to the cylinder geomety (see the pattern in real-space correlations in Fig.~\ref{fig:dmrg_corr}) but it is dominated by short-distance properties and is shown to extrapolate to the same value than $m^2(\pi,0)$. Hence we choose this latter quantity as order parameter for the stripe phase. 

\section{Scaling of the iPEPS variational energies vs inverse environment bond dimension}
\label{app1}

\begin{figure}
\centering
	\subfloat[$K_1 = 0$]{
	\includegraphics[width=55mm, height=50mm]{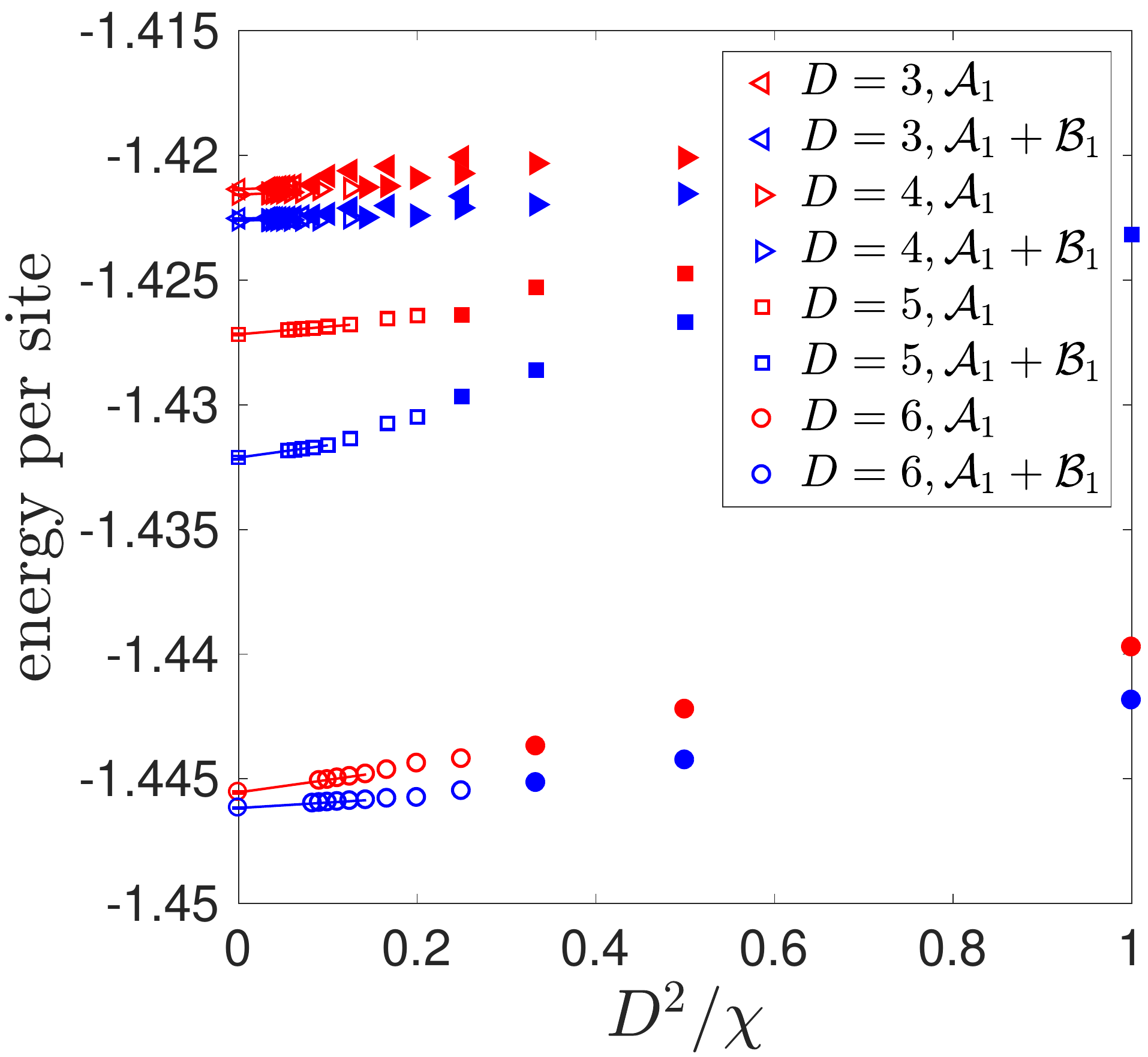}}\\
	\subfloat[$K_1 = 0.05$]{
	\includegraphics[width=56mm, height=50mm]{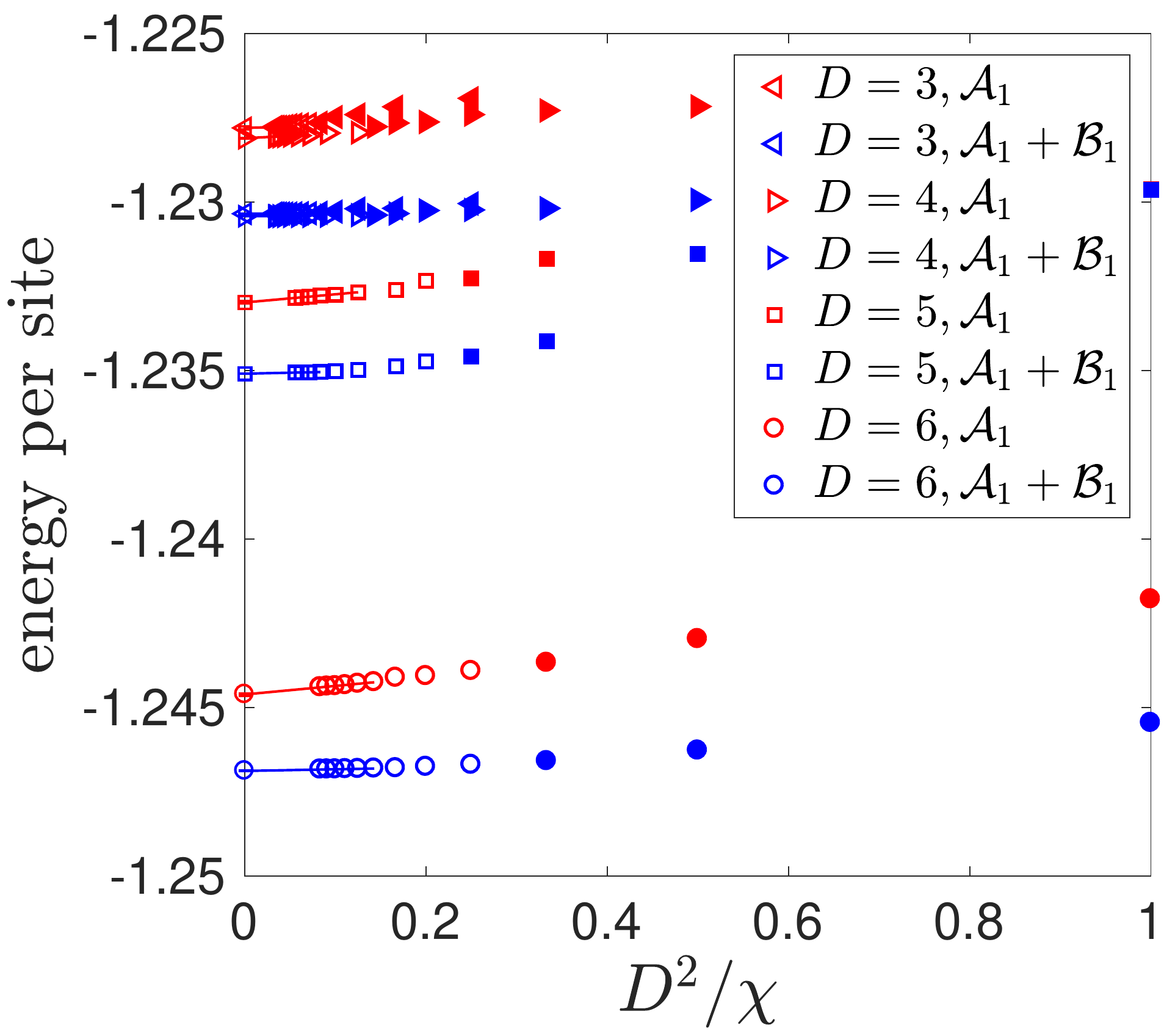}}\\
	\subfloat[$K_1 = 0.1$]{
	\includegraphics[width=56mm, height=50mm]{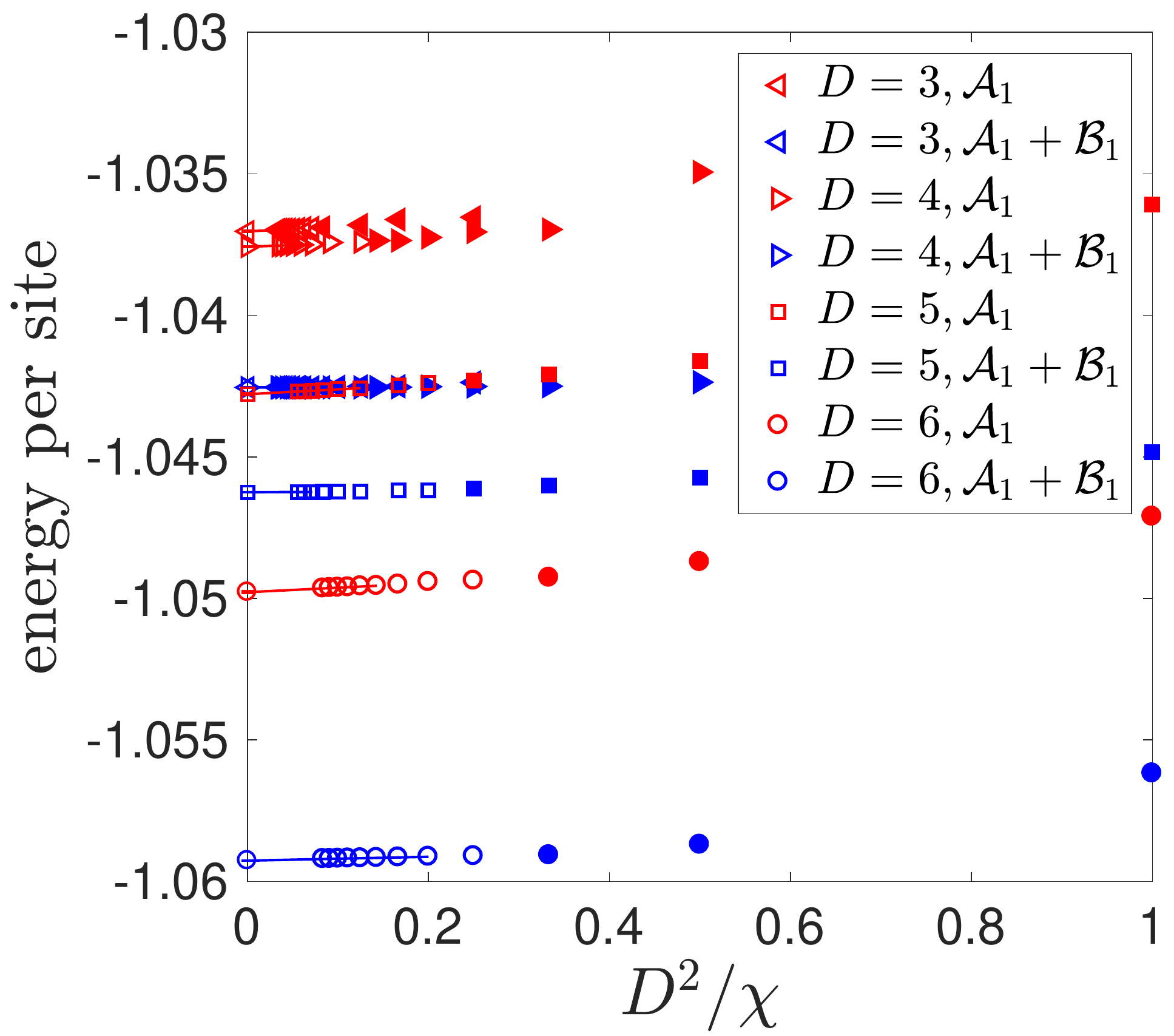}}
\caption{
[Color online] iPEPS energy density for various bond dimension $D$, plotted as a function of $D^2/\chi$.
(a) $K_1 = 0$. (b) $K_1 = 0.05$. (c) $K_1 = 0.1$.
}
\label{fig:PEPS_energy}
\end{figure}

PEPS variational energies are computed directly in thermodynamic limit using an approximate
contraction scheme based on the CTM RG to obtain an effective environment with finite bond dimension $\chi$.
The optimizations with respect to the tensor parameters $\{\lambda_a,\mu_b\}$ ($\mathcal{A}_1 + \mathcal{B}_1$ classes) or $\{\lambda_a\}$ ($\mathcal{A}_1$ classes) have been performed up to $\chi=\chi_{\mathrm{opt}}=108, 112, 100, 108$ for $D=3, 4, 5, 6$, respectively (see main text). For larger $\chi$ a ``frozen'' ansatz obtained with $\chi=\chi_{\rm opt}$ is used. 
Eventually an extrapolation to $\chi=\infty$ is done by fitting the data plotted versus $D^2/\chi$, as shown in
Fig.~\ref{fig:PEPS_energy}.

Also, as it is clear in Fig.~\ref{fig:PEPS_energy},
as we increase $\chi=kD^2$ ($k\in\mathbb{N}_{+}$), the energy goes down continuously.
It is likely that this monotonous behavior applies to arbitrary large $\chi$ (although one cannot prove it), in which case the iPEPS energy can be safely considered as {\it variational}.

\section{Scaling of the maximal correlation length vs environment bond dimension}
\label{app2}

\begin{figure}
\centering
	\subfloat[$D=3$]{
	\includegraphics[width=41.5mm, height=42mm]{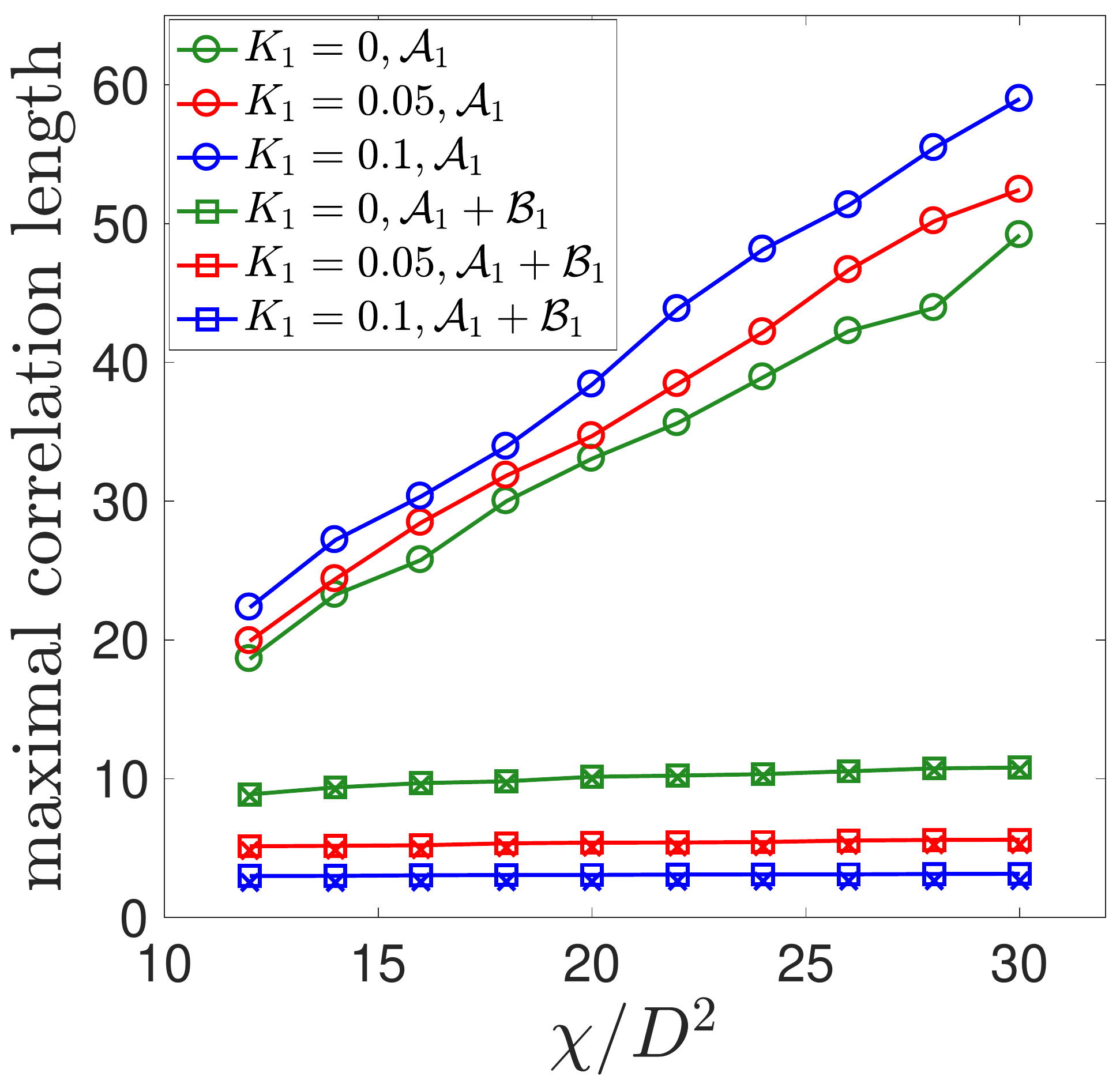}}
	\subfloat[$D=4$]{
	\includegraphics[width=43.5mm, height=42mm]{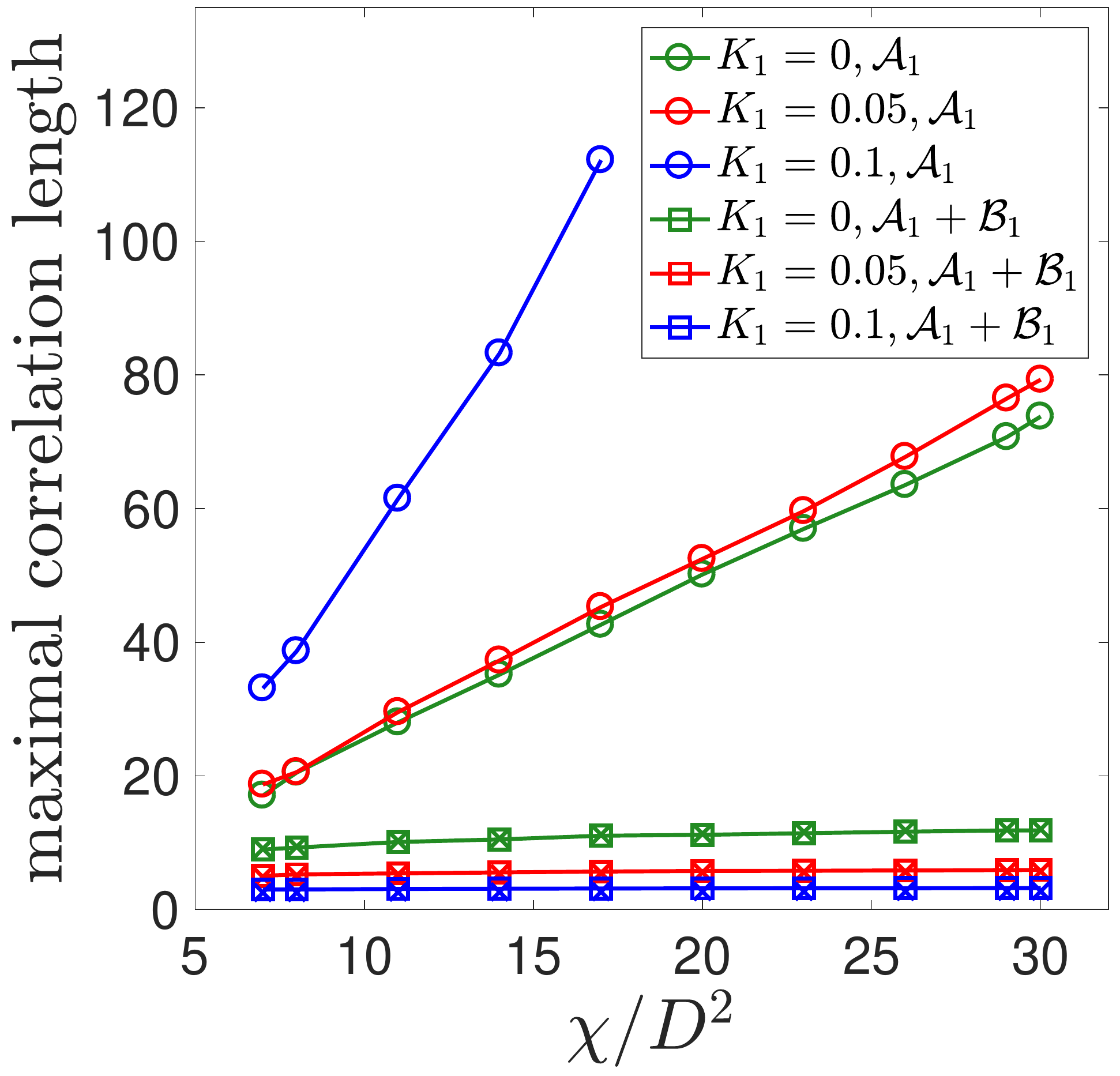}}\\
	\subfloat[$D=5$]{
	\includegraphics[width=42.5mm, height=42mm]{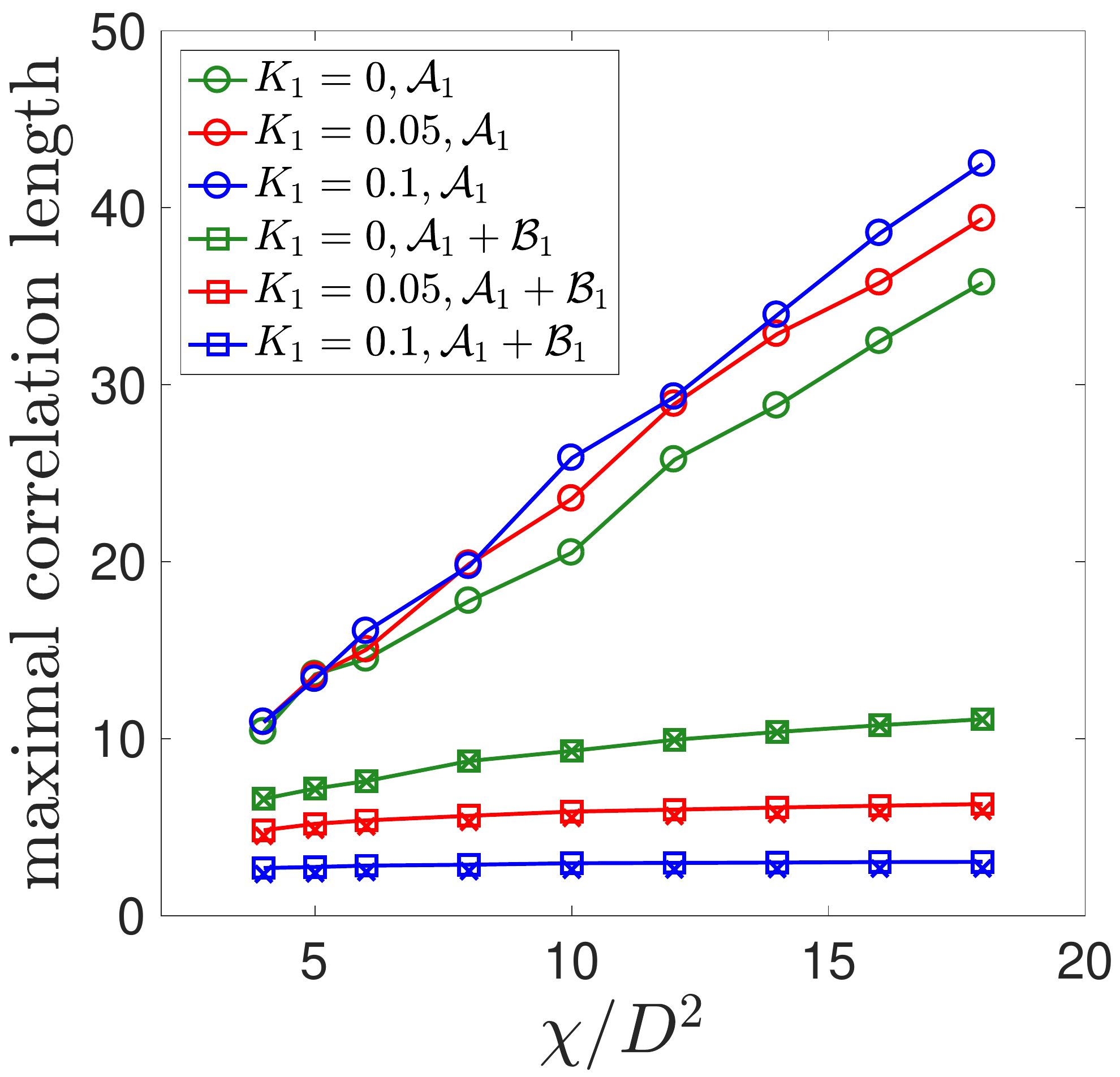}}
	\subfloat[$D=6$]{
	\includegraphics[width=42.5mm, height=42mm]{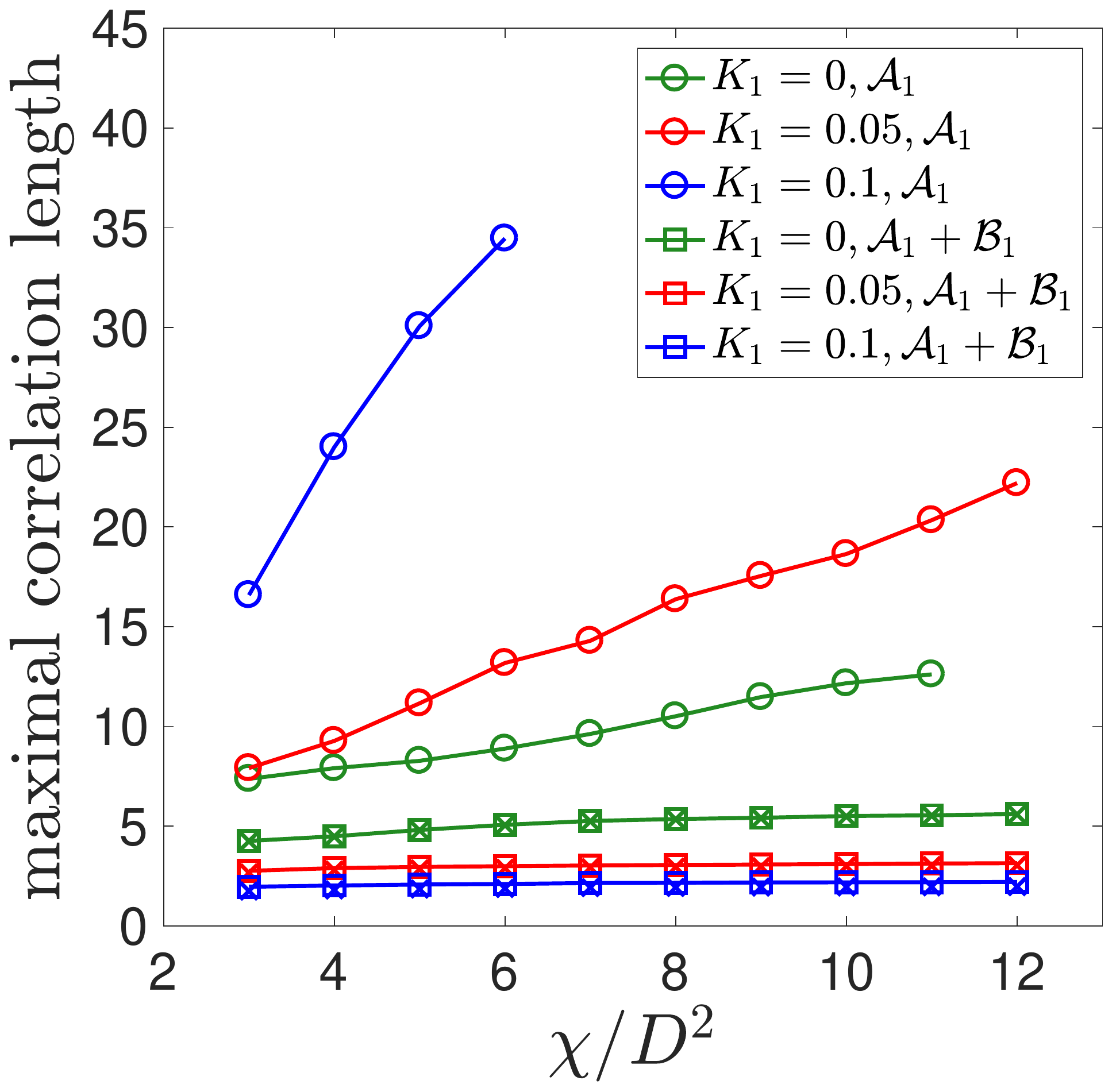}}\\
\caption{
[Color online] iPEPS correlation length $\xi_{\mathrm{max}}$ versus $\chi/D^2$. Each panels correspond to
a given value of the virtual dimension $D$. Open squares / crosses with same color correspond to the weak / strong bond directions of the nematic $\mathcal{A}_1+\mathcal{B}_1$ PEPS. Open circles correspond to the symmetric $\mathcal{A}_1$ PEPS.
}
\label{fig:PEPS_maxCorrLength_bis}
\end{figure}

Here, we present additional data of the maximal correlation length $\xi_{\mathrm{max}}$, shown in Fig.~\ref{fig:PEPS_maxCorrLength_bis}.

The trends that (i) in the symmetric $\mathcal{A}_1$ PEPS, $\xi_{\mathrm{max}}$ diverges faster with increasing $K_1$ while (ii) in the nematic $\mathcal{A}_1+\mathcal{B}_1$ PEPS,
$\xi_{\mathrm{max}}$ converges to smaller and smaller values, can be clearly seen in Fig.~\ref{fig:PEPS_maxCorrLength_bis}. (ii) shows full consistency with a discrete GL symmetry breaking scenario.

\end{document}